\documentclass[compsoc]{IEEEtran}
\usepackage{graphicx}
\usepackage{booktabs}
\usepackage{multirow}
\usepackage[flushleft]{threeparttable}
\usepackage{rotating}
\usepackage{tcolorbox}
\usepackage[numbers]{natbib}
\usepackage[colorlinks, allcolors=black]{hyperref}
\usepackage{cleveref}
\usepackage[htt]{hyphenat}
\usepackage{xspace}

\graphicspath{{figures/}}
\tcbset{arc=2pt, boxrule=\heavyrulewidth, colback=gray!5, colframe=black, boxsep=0pt}

\newcommand{\header}[1]{\smallskip\noindent\textbf{#1}}

\newcommand{\rqi}{How well can we predict the first response latency of maintainers?\xspace}
\newcommand{\rqii}{What are the major predictors of the first response latency of maintainers?\xspace}
\newcommand{\rqiii}{How well can we predict the first response latency of contributors?\xspace}
\newcommand{\rqiv}{What are the major predictors of the first response latency of contributors?\xspace}

\begin{document}

\title{Predicting the First Response Latency of Maintainers and Contributors in Pull Requests}

\author{SayedHassan Khatoonabadi, Ahmad Abdellatif, Diego Elias Costa, and Emad Shihab, \IEEEmembership{Senior Member, IEEE} \IEEEcompsocitemizethanks{\IEEEcompsocthanksitem S. Khatoonabadi, D.E. Costa, and E. Shihab are with the Department of Computer Science \& Software Engineering, Concordia University, Montreal, QC, Canada. E-mail: \{sayedhassan.khatoonabadi, diego.costa, emad.shihab\}@concordia.ca \IEEEcompsocthanksitem A. Abdellatif is with the Department of Electrical \& Software Engineering, University of Calgary, Calgary, AB, Canada. E-mail: ahmad.abdellatif@ucalgary.ca}}

\markboth{IEEE Transactions on Software Engineering}{}

\IEEEtitleabstractindextext{
    \begin{abstract}
    The success of a Pull Request (PR) depends on the responsiveness of the maintainers and the contributor during the review process. Being aware of the expected waiting times can lead to better interactions and managed expectations for both the maintainers and the contributor. In this paper, we propose a machine-learning approach to predict the first response latency of the maintainers following the submission of a PR, and the first response latency of the contributor after receiving the first response from the maintainers. We curate a dataset of 20 large and popular open-source projects on GitHub and extract 21 features to characterize projects, contributors, PRs, and review processes. Using these features, we then evaluate seven types of classifiers to identify the best-performing models. We also conduct permutation feature importance and SHAP analyses to understand the importance and the impact of different features on the predicted response latencies. We find that our CatBoost models are the most effective for predicting the first response latencies of both maintainers and contributors. Compared to a dummy classifier that always returns the majority class, these models achieved an average improvement of 29\% in AUC-ROC and 51\% in AUC-PR for maintainers, as well as 39\% in AUC-ROC and 89\% in AUC-PR for contributors across the studied projects. The results indicate that our models can aptly predict the first response latencies using the selected features. We also observe that PRs submitted earlier in the week, containing an average number of commits, and with concise descriptions are more likely to receive faster first responses from the maintainers. Similarly, PRs with a lower first response latency from maintainers, that received the first response of maintainers earlier in the week, and containing an average number of commits tend to receive faster first responses from the contributors. Additionally, contributors with a higher acceptance rate and a history of timely responses in the project are likely to both obtain and provide faster first responses. Moreover, we show the effectiveness of our approach in a cross-project setting. Finally, we discuss key guidelines for maintainers, contributors, and researchers to help facilitate the PR review process.
    \end{abstract}
    
}

\maketitle

\IEEEraisesectionheading{\section{Introduction}}
\IEEEPARstart{P}{ull}-based development has become a common paradigm for contributing to and reviewing code changes in numerous open-source projects \citep{gousios_exploratory_2014, zhu_effectiveness_2016}. Pull Requests (PRs) are the driving force behind the maintenance and evolution of these projects, encompassing everything from bug fixes to new features. Contributors initiate this collaborative process by submitting a PR that proposes changes for integration into the project. The PR then undergoes a review process, during which the contributor revises the changes based on feedback from the project maintainers. This cycle repeats until the PR satisfies the maintainers' requirements for getting merged \citep{jiang_how_2022, li_opportunities_2022}.

The success of the PR depends on the responsiveness of both the maintainers and the contributor during the review process \citep{khatoonabadi_wasted_2023, li_are_2022, gousios_work_2016, gousios_work_2015}. Timely responses from the maintainers set a positive tone for the entire review process, increasing the likelihood of the contributor continuing the review process towards completion \citep{khatoonabadi_wasted_2023, li_opportunities_2022}. Conversely, delayed responses are often perceived as negligence, increasing the risk of the contributor abandoning the PR \citep{khatoonabadi_wasted_2023, li_are_2022, wang_why_2019}. Once the maintainers have responded, the contributor's promptness in addressing the feedback is equally crucial. Timely responses help maintain the momentum of the review process, whereas delayed responses can cause it to stale \citep{khatoonabadi_wasted_2023, li_opportunities_2022, khatoonabadi_understanding_2023}.

Knowing the expected waiting times can lead to better interactions and managed expectations for both sides. Contributors, when aware of anticipated waiting times, can adjust their schedules accordingly, reducing uncertainty and preserving their motivation throughout the review process \citep{khatoonabadi_wasted_2023, gousios_work_2016}. Maintainers, aware of possible delays in contributor responses, can proactively offer additional support or take action to mitigate potential blockers \citep{khatoonabadi_wasted_2023}. This awareness also allows maintainers to better allocate their time and resources and prioritize PR reviews \citep{gousios_work_2015}. Furthermore, analyzing response time trends can help projects pinpoint and rectify bottlenecks, thereby enhancing the efficiency and effectiveness of their PR review workflows.

The first responses are of particular importance as they not only directly influence the duration \citep{hasan_understanding_2023, zhang_pull_2022} and the outcome \citep{zhang_pull_2023, khatoonabadi_wasted_2023, li_are_2022} of the review process, but also the likelihood of future contributions by the contributor \citep{assavakamhaenghan_does_2023, hasan_understanding_2023}. Despite the critical role of first responses, existing approaches only aim to understand the characteristics of PRs with a longer time to first bot or human response \citep{hasan_understanding_2023}, predict the completion time of PRs \citep{de_lima_junior_predicting_2021, maddila_predicting_2019}, or devise approaches to nudge overdue PRs \citep{maddila_nudge_2023, shan_using_2022}. Our study bridges this gap by proposing a machine learning approach to predict: (1) the first response latency of the \textit{maintainers} following the submission of a PR, and (2) the first response latency of the \textit{contributor} after receiving the first response from the maintainers.

For this purpose, we start by curating a dataset of 20 popular and large open-source projects on GitHub. Next, we extract 21 features to characterize projects, contributors, PRs, and review processes. Using these features, we then evaluate seven types of classifiers to identify the best-performing models. Finally, we perform permutation feature importance \citep{fisher_all_2019} and SHAP \citep{lundberg_unified_2017} analyses to understand the importance and impact of different features on the predicted response latencies. In summary, we aim to answer the following four research questions:

\begin{itemize}
    \item[\textbf{RQ1:}] \textbf{(Maintainers) \rqi} We find that the CatBoost models outperform other models in predicting the first response latency of maintainers, achieving an average improvement of 29\% in AUC-ROC and 51\% in AUC-PR compared to a dummy classifier that always returns the majority class across the studied projects.
    \item[\textbf{RQ2:}] \textbf{(Maintainers) \rqii} We find that PRs submitted earlier in the week, containing an average or slightly above-average number of commits at submission, and with more concise descriptions are more likely to get faster responses. Similarly, contributors with a higher acceptance rate and a history of timely responses in the project tend to obtain quicker responses.
    \item[\textbf{RQ3:}] \textbf{(Contributors) \rqiii} Similar to the first response latency of maintainers, we find that the CatBoost models outperform other models in predicting the first response latency of contributors, achieving an average improvement of 39\% in AUC-ROC and 89\% in AUC-PR compared to a dummy classifier that always returns the majority class across the studied projects.
    \item[\textbf{RQ4:}] \textbf{(Contributors) \rqiv} We find that contributors of PRs that experienced a lower first response latency from maintainers, PRs that received the first response of maintainers earlier in the week, and PRs containing an average or slightly above-average number of commits till the first response of maintainers are more likely to provide faster responses. Similarly, contributors with a history of timely responses in the project and with a higher acceptance rate tend to give quicker responses.
\end{itemize}

Finally, we evaluate our approach in a cross-project setting. This is especially useful for new projects with limited historical data to build accurate models. Compared to a dummy classifier that always returns the majority class, the models achieve an average improvement of 33\% in AUC-ROC and 58\% in AUC-PR for maintainers, as well as an average improvement of 42\% in AUC-ROC and 95\% in AUC-PR for contributors. Furthermore, we find that the key predictors in the cross-project setting are: submission day, number of commits, contributor acceptance rate, historical maintainers responsiveness, and historical contributor responsiveness for maintainers' first response latency; and first review latency, review day, historical contributor responsiveness, number of commits, and contributor activity within the PR for contributors' first response latency. Finally, based on our findings, we discuss key guidelines for maintainers, contributors, and researchers to help facilitate the PR review process. We expect our approach to enhance collaboration between maintainers and contributors by enabling them to anticipate waiting times and take proactive actions to mitigate potential challenges during the PR review process.

\smallskip
\header{Our Contributions.}
In summary, we make the following contributions in this paper:

\begin{itemize}
    \item To the best of our knowledge, we are the first to propose machine learning models for classifying the first response latency of maintainers and contributors.
    \item We investigate the major predictors of the first response latency of maintainers and contributors and discuss the impact of the features on the anticipated waiting periods.
    \item To promote the reproducibility of our study and facilitate future research on this topic, we publicly share our scripts and dataset online at \url{https://doi.org/10.5281/zenodo.10119283}.
\end{itemize}

\header{Paper Organization.}
The rest of this paper is organized as follows. \Cref{sec:study_design} overviews the design of our study. \Cref{sec:maintainers,sec:contributors} report the results for predicting the first response latency of maintainers and contributors, respectively. \Cref{sec:cross-project} evaluates our approach in a cross-project setting and \Cref{sec:implications} discusses the implications of our findings. \Cref{sec:limitations} discusses the limitations of our study and \Cref{sec:related_work} reviews the related work. Finally, \Cref{sec:conclusion} concludes the paper.

\section{Study Design}
\label{sec:study_design}
The main objective of this study is to develop machine learning models for predicting the first response latency of the maintainers following the submission of a PR; as well as the first response latency of the contributor after receiving the first response from the maintainers. Additionally, we aim to identify and discuss the impact of the key features that significantly influence the predicted first response latency of maintainers and contributors. In the following, we explain the methodology and design of our study in detail.

\subsection{Studied Projects}
To ensure that we have enough historical data for our study, we seek popular open-source projects with the largest histories of pull-based development. For this purpose, we rely on GitHub as a pioneer in supporting pull-based development and the largest open-source ecosystem \citep{github_state_2022}, which has also been the subject of numerous code review studies \citep{badampudi_modern_2023, davila_systematic_2021}. To identify such projects, we use the number of stars as a proxy for the popularity of the projects \citep{borges_whats_2018} and retrieve the list of the top 1,000 most-starred projects. Among these projects, we then select the top 20 with the highest number of PRs. \Cref{tab:projects} provides an overview of the projects that we selected for our case study. In summary, the selected projects have thousands of PRs (median of 42,630), thousands of stars (median of 52,226), hundreds of contributors (median of 3,240), tens of maintainers (median of 201), and years of pull-based development history (median of 106 months). Additionally, these projects span multiple application domains and programming languages, providing a more diverse selection of projects for our case study. We collected the timeline of activities for the selected projects on December 1st, 2022. The timeline of activities of PRs is provided by the GitHub API \citep{github_rest_2023} and includes the details (e.g., type, actor, and time) of all the events (e.g., commits, comments, and resolutions) that occurred during the lifecycle of a PR \citep{github_timeline_2023, github_pulls_2023, github_issues_2023}.

\begin{table*}
    \centering
    \caption{Overview of the projects selected for our study.}
    \label{tab:projects}
    \resizebox{\textwidth}{!}{%
        \begin{tabular}{@{}l|ccccc|ll@{}}
            \toprule
            \textbf{Project} & \textbf{PRs} & \textbf{Stars} & \textbf{Contributors} & \textbf{Maintainers} & \textbf{Age (Months)} & \textbf{Domain}                    & \textbf{Language} \\
            \midrule
            Odoo             & 89,513       & 27,267         & 2,678                 & 237                  & 102                   & Business Management System         & JavaScript        \\
            Kubernetes       & 72,001       & 94,103         & 5,613                 & 361                  & 101                   & Container Orchestration System     & Go                \\
            Elasticsearch    & 60,582       & 61,986         & 3,016                 & 299                  & 153                   & Data Analytics Engine              & Java              \\
            PyTorch          & 60,072       & 60,605         & 3,610                 & 350                  & 75                    & Machine Learning Framework         & C++               \\
            Rust             & 58,563       & 74,949         & 4,566                 & 135                  & 149                   & Programming Language               & Rust              \\
            DefinitelyTyped  & 53,843       & 41,717         & 18,980                & 548                  & 121                   & TypeScript Type Definitions        & TypeScript        \\
            HomeAssistant    & 48,478       & 56,269         & 4,646                 & 913                  & 110                   & Home Automation System             & Python            \\
            Ansible          & 48,262       & 55,575         & 7,835                 & 93                   & 128                   & IT Automation Platform             & Python            \\
            CockroachDB      & 45,884       & 26,127         & 725                   & 203                  & 105                   & Database Management System         & Go                \\
            Swift            & 45,560       & 61,197         & 1,333                 & 230                  & 85                    & Programming Language               & C++               \\
            Flutter          & 39,700       & 146,697        & 2,287                 & 236                  & 92                    & Software Development Kit           & Dart              \\
            Spark            & 38,809       & 34,459         & 3,463                 & 80                   & 105                   & Data Analytics Engine              & Scala             \\
            Python           & 37,792       & 49,172         & 3,779                 & 104                  & 69                    & Programming Language               & Python            \\
            Sentry           & 35,489       & 32,665         & 834                   & 170                  & 146                   & Application Performance Monitoring & Python            \\
            PaddlePaddle     & 32,738       & 19,230         & 831                   & 500                  & 75                    & Machine Learning Framework         & C++               \\
            Godot            & 30,886       & 55,580         & 2,688                 & 63                   & 106                   & Game Engine                        & C++               \\
            Rails            & 30,386       & 51,852         & 6,435                 & 114                  & 175                   & Web Application Framework          & Ruby              \\
            Grafana          & 30,373       & 52,599         & 2,657                 & 199                  & 107                   & Data Visualization Platform        & TypeScript        \\
            ClickHouse       & 29,820       & 26,262         & 1,377                 & 54                   & 77                    & Database Management System         & C++               \\
            Symfony          & 29,009       & 27,679         & 4,240                 & 51                   & 154                   & Web Application Framework          & PHP               \\
            \bottomrule
        \end{tabular}
    }
\end{table*}

\subsection{Identification of First Responses}
In this study, we focus on responses from project maintainers, as their deeper knowledge of the project enables them to evaluate PRs more effectively. Following an approach similar to \citet{bock_automatic_2023}, we consider developers with privileged access within the project PRs as PR maintainers. This status not only enables them to perform key PR maintenance tasks, but also allows them to make decisions about accepting or rejecting PRs. To identify PR maintainers, we analyze each developer's activity within the project PRs, looking for any of the following privileged PR-related events \citep{bock_automatic_2023}: \texttt{added\_to\_project}, \texttt{deployed}, \texttt{deployment\_environment\_changed}, \texttt{locked}, \texttt{merged}, \texttt{moved\_columns\_in\_project}, \texttt{removed\_from\_project}, \texttt{review\_dismissed}, \texttt{unlocked}, and \texttt{user\_blocked}. In addition to these events, we consider actions such as merging a PR using other nonstandard methods (e.g., through commit messages that include keywords to resolve a PR, like ``resolves \#123'') or closing someone else's PR as privileged events (since users can close their own PRs without requiring any special access). Finally, we classify any developer associated with any of the aforementioned actions as a PR maintainer of the project henceforth.

Similar to \citep{hasan_understanding_2023}, we define the first response of maintainers as the first feedback (i.e., \texttt{commented}, \texttt{reviewed}, \texttt{line-commented}, and \texttt{commit-commented}) or resolution (i.e., \texttt{merged}, \texttt{closed}, and \texttt{reopened}) by a maintainer (excluding bots) other than the contributor after a PR is submitted. We also define the first response of contributors as the first update (i.e., \texttt{committed} and \texttt{head\_ref\_force\_pushed}), feedback (i.e., \texttt{commented}, \texttt{reviewed}, \texttt{line-commented}, and \texttt{commit-commented}, or resolution (i.e., \texttt{closed} and \texttt{reopened}) by the contributor after receiving the first response from a maintainer. We identified bots by marking actors listed in any of the three ground-truth datasets \citep{abdellatif_bothunter_2022, wang_specialized_2022, golzadeh_ground-truth_2021}, as well as those with names ending in \texttt{bot} or \texttt{[bot]}. Additionally, we manually inspected actors with high activity levels or fast response times to detect any potential bots that may have been overlooked.

\subsection{Feature Extraction}
To train machine learning models for predicting the first response latency of either maintainers or contributors, we need to extract a set of relevant features that can potentially be predictive of their first response latencies. For this purpose, we consult the literature on pull-based development \citep{zhang_pull_2022, zhang_pull_2023} and also draw from our previous experience studying PR abandonment \citep{khatoonabadi_wasted_2023, khatoonabadi_understanding_2023}. As outlined in \Cref{tab:features}, we extract a total of 21 features covering four different dimensions: (i) project characteristics, (ii) contributor characteristics, (iii) PR characteristics, and (iv) review process characteristics. Features in the table are denoted by `M', `C', or `MC', indicating their use in the models for predicting the response latency of maintainers, contributors, or both, respectively. The features are measured using the data available at different time points: features for predicting the maintainer response latency are measured at the submission time of PR, while features for predicting the contributor response latency are measured at the time when the PR receives its first response from a maintainer. In addition, all measurements are based on the UTC timezone as directly provided by the GitHub API \citep{github_rest_2023}. Also, project features (except Project Backlog) are measured over the last three months, similar to \citep{zhang_pull_2022, zhang_pull_2023}, to better reflect recent fluctuations as a project matures. Our criteria for selecting the features are as follows:
\begin{itemize}
    \item \textbf{Conceptual Availability:} The feature should be conceptually available at the time of interest. For predicting the maintainer's first response latency, the feature should be available at the submission time of a PR; and for predicting the contributor's first response latency, the feature should be available at the first response time of maintainers. For example, we did not include the CI execution results for predicting the maintainer's first response latency as they are not available at the submission time of a PR.
    \item \textbf{Universal Availability:} The feature should be available for all PRs of the studied projects to ensure a consistent set of features across the projects and also to avoid requiring imputation techniques to fill in the missing values. This is because any imputation technique inherently alters the distribution of data and thus will reduce the interpretability of our results. For example, we did not include the CI execution results for predicting the contributor's first response latency as they are only available for some PRs and for some projects.
    \item \textbf{Feature Measurability:} The feature should be measurable using the data available through the GitHub API \citep{github_rest_2023}. For example, we did not include the number of followers because the GitHub API does not reveal the number of followers of a user at a historical point in time and only provides the latest count.
    \item \textbf{Feature Non-Redundancy:} The feature should not be just another operationalization of the same concept already included in our set of features. In cases where several features can be combined into one feature, we preferred to include the more generic feature. For example, we include the number of changed lines instead of including the number of added lines and the number of deleted lines as two separate features.
    \item \textbf{Measurement Accuracy:} The measurement of the feature should be accurate and should not rely on ML models or heuristics. This is because inaccurate values hinder the interpretability of the feature. For example, we did not include the existence of @tags as we frequently observed that @tags are used for purposes other than tagging a user.
\end{itemize}

In the following, we explain the relevance and measurement of each feature in detail.

\begin{table*}
    \centering
    \caption{Features extracted to predict the first response latency of maintainers (M) and contributors (C).}
    \label{tab:features}
    \resizebox{\textwidth}{!}{%
        \begin{tabular}{@{}l|ll|c@{}}
            \toprule
            \textbf{Dimension}                       & \textbf{Feature}           & \textbf{Description}                                                     & \textbf{Model} \\
            \midrule
            \multirow{5}{*}{\textbf{Project}}        & Submission Volume          & Number of submitted PRs to the project over the last 3 months            & MC             \\
                                                     & Project Backlog            & Number of unresolved PRs in the project                                  & MC             \\
                                                     & Maintainers Availability   & Number of active maintainers in the project over the last 3 months       & MC             \\
                                                     & Maintainers Responsiveness & Median first response latency of the maintainers over the last 3 months  & MC             \\
                                                     & Community Size             & Number of active community members in the project over the last 3 months & MC             \\
            \midrule
            \multirow{4}{*}{\textbf{Contributor}}    & Contributor Experience     & Number of prior PRs by the contributor                                   & MC             \\
                                                     & Contributor Performance    & Ratio of the merged PRs of the contributor                               & MC             \\
                                                     & Contributor Backlog        & Number of unresolved PRs by the contributor                              & MC             \\
                                                     & Contributor Responsiveness & Median first response latency of the contributor in prior PRs            & MC             \\
            \midrule
            \multirow{6}{*}{\textbf{Pull Request}}   & Description Length         & Number of words in the title and description of the PR                   & MC             \\
                                                     & Commits                    & Number of commits in the PR                                              & MC             \\
                                                     & Changed Lines              & Number of changed lines in the PR                                        & MC             \\
                                                     & Changed Files              & Number of changed files in the PR                                        & MC             \\
                                                     & Submission Day             & Weekday of the submission time of the PR                                 & M              \\
                                                     & Submission Hour            & Hour of the submission time of the PR                                    & M              \\
            \midrule
            \multirow{6}{*}{\textbf{Review Process}} & Review Day                 & Weekday of the first response of the maintainer                          & C              \\
                                                     & Review Hour                & Hour of the first response of the maintainer                             & C              \\
                                                     & Review Latency             & First response latency of the maintainer in the PR                       & C              \\
                                                     & Contributor Activity       & Number of events by the contributor in the PR                            & C              \\
                                                     & Participants Activity      & Number of events by the participants in the PR                           & C              \\
                                                     & Bots Activity              & Number of events by the bots in the PR                                   & C              \\
            \bottomrule
        \end{tabular}
    }
\end{table*}

\header{Submission Volume.} Increased PR submissions can overwhelm maintainers, resulting in delayed response times \citep{gousios_work_2015}. However, a high volume of submissions indicates an active project, which can attract contributors and positively influence their responsiveness. To quantify this feature, we count the number of PRs submitted to the project over the last three months.

\header{Project Backlog.} A sizeable backlog of unresolved PRs could overwhelm maintainers, potentially resulting in extended response times \citep{hasan_understanding_2023, zhang_pull_2022}. Furthermore, contributors may perceive a substantial backlog as a sign of inattentiveness, which may discourage them and adversely affect their willingness to respond promptly. To quantify this feature, we count the number of unresolved PRs in the project.

\header{Maintainers Availability.} More available maintainers facilitate efficient workload distribution, potentially leading to quicker response times \citep{zhang_pull_2022, gousios_work_2015}. Higher availability of maintainers indicates an actively maintained and supportive project, boosting contributors' confidence and responsiveness. To quantify this feature, we count the number of active maintainers in the project over the last three months.

\header{Maintainers Responsiveness.} The past responsiveness of maintainers can serve as an indicator of their future response times. Specifically, a history of delayed responses may be interpreted as inattentiveness or lack of engagement, which can demotivate contributors, subsequently dampening their responsiveness \citep{khatoonabadi_wasted_2023, li_are_2022}. To quantify this feature, we calculate the median first response latency of maintainers for PRs they responded to over the last three months.

\header{Community Size.} High levels of community engagement can introduce diverse inputs and increased demands on maintainers, potentially leading to delayed responses \citep{zhang_pull_2022, gousios_work_2015}. Nevertheless, active community participation reflects a thriving project, which can encourage prompt responses by contributors. To quantify this feature, we count the number of active practitioners (excluding maintainers and bots) in the project over the last three months.

\header{Contributor Experience.} Experienced contributors tend to submit higher-quality PRs and communicate more effectively, potentially expediting responses from maintainers \citep{hasan_understanding_2023, zhang_pull_2022}. Their familiarity with the project's dynamics and expectations often leads to quicker responses to feedback and revision requests \citep{gousios_work_2016}. To quantify this feature, we calculate the number of PRs a contributor has previously submitted to the project.

\header{Contributor Performance.} Contributors who consistently have a high success rate with their submissions often produce PRs that align closely with the project's standards, likely receiving quicker responses from maintainers \citep{hasan_understanding_2023, zhang_pull_2022}. On the other hand, contributors who are familiar with the project's expectations typically respond more promptly to feedback \citep{gousios_work_2016}. To quantify this feature, we calculate the ratio of a contributor's successfully merged PRs to their total submissions in the project.

\header{Contributor Backlog.} A backlog of PRs from a contributor may suggest they are overextended or tend to submit PRs that require extensive review, potentially leading to delayed responses from maintainers \citep{khatoonabadi_wasted_2023}. Additionally, having multiple pending submissions may divide the contributor's attention, slowing their response times. To quantify this feature, we count the number of a contributor's unresolved PRs in the project.

\header{Contributor Responsiveness.} The past responsiveness of a contributor can serve as an indicator of their future response times. Previous timely responses not only exhibit commitment but also foster a cycle of timely feedback from maintainers \citep{khatoonabadi_wasted_2023, li_are_2022}. To quantify this feature, we calculate the median latency of the contributor's first responses in prior PRs within the project.

\header{Description Length.} A description that is both detailed and concise can significantly aid maintainers in evaluating the proposed changes, potentially resulting in faster response times \citep{hasan_understanding_2023, zhang_pull_2022}. A clear description may also minimize the need for further clarification or modifications, allowing contributors to promptly and efficiently address feedback from maintainers \citep{gousios_work_2016}. To quantify this feature, we count the number of words in both the title and description of the PR.

\header{Commits.} PRs with fewer commits often facilitate a smoother review process, leading to quicker responses from maintainers \citep{hasan_understanding_2023, zhang_pull_2022}. On the other hand, contributors who craft meaningful commits are also likely more motivated and attentive to feedback, resulting in prompt responses \citep{gousios_work_2016}. To quantify this feature, we count the number of commits submitted to the PR.

\header{Changed Lines.} PRs with extensive changes are often more difficult and time-consuming for maintainers to review, leading to delayed responses \citep{hasan_understanding_2023, zhang_pull_2022}. Contributors of such PRs also typically need more time to address the changes requested by the maintainers \citep{gousios_work_2016}. To quantify this feature, we count the number of lines that have been changed in the commits included in the PR.

\header{Changed Files.} PRs that touch multiple files require the reviewer to go through changes across different files, which could extend the time it takes for maintainers to respond \citep{hasan_understanding_2023, zhang_pull_2022}. Similar to PRs with many changed lines, those that change many files often reflect a significant effort by contributors, making them more eager to respond to feedback. However, addressing feedback on multiple files may take contributors more time \citep{gousios_work_2016}. To quantify this feature, we count the number of files changed in the PR's commits.

\header{Submission Day.} PRs submitted closer to weekends tend to receive slower responses due to reduced activity from maintainers compared to normal working days \citep{thongtanunam_review_2017, gousios_work_2015}. To quantify this feature, we record the day of the week on which the PR is submitted.

\header{Submission Hour.} Similar to submission day, PRs submitted during the usual working or active hours of maintainers tend to receive quicker responses. In contrast, those submitted outside of these hours might face delays, as maintainers may not be readily available or active \citep{gousios_work_2015}. To quantify this feature, we record the hour on which the PR is submitted.

\header{Review Day.} Similar to PR submission day, reviews conducted on weekdays might align more closely with the schedules of contributors who approach their work on the project professionally, encouraging timely responses \citep{gousios_work_2015}. To quantify this feature, we record the day of the week when the maintainer's first response is submitted.

\header{Review Hour.} If the review hour aligns with contributors' usual active hours, it improves the chances that the contributor respond more promptly \citep{gousios_work_2015}. To quantify this feature, we record the hour of the maintainer's first response to the PR.

\header{Review Latency.} Prior studies show that the first response latency of maintainers is directly correlated with the total duration of the PR review process \citep{hasan_understanding_2023, zhang_pull_2022}. Timely responses from maintainers enhance contributors' engagement with the project \citep{gousios_work_2016}. However, delays in reviews can lead to frustration and potential PR abandonment \citep{khatoonabadi_wasted_2023, li_are_2022, wang_why_2019}. To quantify this feature, we measure the time it takes for maintainers to issue their first response to a PR in hours.

\header{Contributor Activity.} The level of activity exhibited by a contributor during the review process can be a sign of their engagement and dedication to the PR, which often correlates with quicker responsiveness to feedback \citep{li_are_2022}. To quantify this feature, we count the number of events initiated by the contributor following the PR's submission.

\header{Participants Activity.} The level of activity from participants, excluding maintainers and bots, during the review process indicates the community's interest and engagement with a specific PR. When a PR attracts attention and constructive comments from the community, it tends to encourage and motivate contributors to respond more quickly \citep{li_are_2022}. To quantify this feature, we count the number of events initiated by these participants following the PR's submission.

\header{Bots Activity.} Bots play an essential role in facilitating the review process by automating repetitive tasks. However, excessive or inappropriate bot activity disrupts contributors, potentially impeding their responsiveness and engagement \citep{wessel_bots_2022, wessel_quality_2022, khatoonabadi_understanding_2023}. To quantify this feature, we count the number of events triggered by bots after the submission of the PR.

\subsection{Data Preprocessing}
In the following, we explain how we preprocess the dataset before constructing our models.

\header{Data Filtering.} We exclude PRs submitted by bots to avoid skewing response latency data. Bots typically receive different treatment from maintainers, leading to unusual response patterns \citep{wyrich_bots_2021}. This can misrepresent the typical response dynamics between human maintainers and contributors as bot interactions do not mirror human behavior.

\header{Feature Correlation Analysis.} The presence of multicollinearity \citep{dormann_collinearity_2013} can adversely affect both the performance and interpretability of machine learning models \citep{jiarpakdee_impact_2021}. To mitigate this issue, we identify highly correlated features using the Spearman rank correlation test \citep{spearman_proof_2010} as a nonparametric test that does not require normally distributed data. Then, we conduct the analysis on the entire dataset to ensure that the identified correlations are consistent across all the studied projects. For strongly correlated features with \(|\rho| \geq 0.6\) (as suggested by \citep{evans_straightforward_1996}), we keep the feature that we believe could be more practical and useful to the stakeholders of our models (i.e., project maintainers and contributors). In the following, we explain the rationale for our selections among each set of strongly correlated features:
\begin{itemize}
    \item \textbf{Submission Volume and Maintainers Availability:} We selected maintainers availability because projects have more direct control over the availability of maintainers compared to the external nature of submission volumes.
    \item \textbf{Contributor Experience and Contributor Performance:} We selected contributor performance (i.e., ratio of the merged PRs of the contributor) because it provides a direct measure of quality over quantity, thereby minimizing potential bias against newer contributors who may have previously submitted fewer PRs but have a high success rate.
    \item \textbf{Commits, Changed Lines, and Changed Files:} We selected the number of commits because it is a more consistent measure of PR complexity and is less prone to extreme values compared to the number of changed lines or files, which can vary significantly depending on the nature of proposed changes.
\end{itemize}

\header{Feature Transformation.} Skewed data can negatively impact machine learning models that expect normally distributed data. To address this issue, we apply log transformation \citep{izenman_modern_2008} to the features using \(\log(x+1)\). For models other than tree-based ones, where having a comparable scale for features is crucial for optimal performance, we further standardize the features using the Z-score normalization \citep{izenman_modern_2008} technique to achieve a distribution with a mean of zero and a standard deviation of one.

\subsection{Model Construction and Evaluation}
We build and validate various machine learning models that use the extracted features for predicting the first response latency of maintainers and contributors. The models classify the response times into one of the following three classes: 1) within 1 day, 2) 1 day to 1 week, 3) more than 1 week. This classification scheme is adapted from the study conducted by \citet{hasan_understanding_2023}, which categorized the response times into four groups: 1) within 1 day, 2) 1 day to 1 week, 3) 1 week to 1 month, and 4) more than 1 month. However, we combined the last two categories since PRs with first responses of more than one month are rare in our dataset (less than 2\% for either of first maintainers and contributors responses in most of our studied projects). The distribution of the first response latency of maintainers and contributors across the studied projects can be found in Appendix.

We experiment with various classifier models to identify which ones most accurately predict the first response latency of maintainers and contributors across the studied projects. The selected models include CatBoost (CB) \citep{prokhorenkova_catboost_2018}, K-Nearest Neighbors (KNN), Logistic Regression (LR), Naive Bayes (NB), Neural Network (NN), Random Forest (RF), and Support Vector Machine (SVM). CatBoost is recommended for multiclass imbalanced datasets \citep{tanha_boosting_2020}. Other models are also popular in the software engineering literature \citep{wang_machinedeep_2023, yang_predictive_2022}. For each project, we train two distinct models using the corresponding selected features: one focusing on the first response latency of maintainers and another on the first response latency of contributors. To evaluate the performance of our models, we use Time Series cross-validation. This technique is a variation of K-Fold cross-validation that takes into account the temporal nature of our data, thus preventing future information from leaking to training sets. In each split, the first \(k\) folds serve as the training set and the \((k+1)\)-th fold serves as the test set. To measure the performance of the models, we rely on the following two threshold-independent metrics that are commonly used to evaluate model performance in the presence of an imbalanced dataset \citep{haibo_he_learning_2009}:

\begin{itemize}
    \item \header{AUC-ROC:} This metric measures the area under the Receiver Operating Characteristic (ROC) curve \citep{bradley_use_1997}, which plots the true positive rate against the false positive rate at different thresholds. AUC-ROC values range from 0 to 1, with a value greater than 0.5 indicating that the model performs better than a no-skill classifier.
    \item \header{AUC-PR:} This metric measures the area under the Precision-Recall (PR) curve \citep{fawcett_introduction_2006}, which plots the precision against the recall across different thresholds. AUC-PR also ranges from 0 to 1. However, the performance of a no-skill classifier is determined by the class distribution.
\end{itemize}

These metrics provide a comprehensive assessment of model discriminative capability across various probability thresholds, allowing for a robust and generalized evaluation of performance. To measure these metrics, we adopt the One-vs-Rest (OvR) approach \citep{rifkin_defense_2004} commonly used for evaluating multiclass classifiers \citep{dellanna_evaluating_2023}. This approach decomposes the multiclass classification problem into multiple binary classification problems, each focusing on a single class against all others. The metrics are then computed for each binary problem and averaged (macro) to provide an aggregated overview of the model's capability across different classes. Using this approach, we can perform a detailed evaluation of the model's capability to distinguish each class, ensuring that the model is not biased towards any particular class and can perform well across all classes. Finally, we calculate the relative improvement compared to a dummy classifier that always returns the majority class as our baseline (as implemented in \texttt{scikit-learn}). To identify the best-performing models, we rely on the nonparametric Scott-Knott ESD test \citep{tantithamthavorn_impact_2019}. This is a multiple comparison approach that leverages hierarchical clustering to partition the set of median values of techniques into statistically distinct groups with non-negligible differences.

\subsection{Model Analysis}
To compare the relative importance of different features, we perform permutation feature importance analysis \citep{fisher_all_2019, rajbahadur_impact_2022} for each project. This approach permutes a feature to break the association between the feature and the outcome (i.e., the response latency). The importance of the feature is then measured by how much error the permutated data introduces compared to the original error (i.e., loss in AUC-ROC in our case). Therefore, the most important features have the largest impact on the performance of our models and thus are more useful for making accurate predictions. After calculating the importance of each feature in each project, we apply the nonparametric Scott-Knott ESD test \citep{tantithamthavorn_impact_2019} to obtain the ranking of each feature. To understand the impact of each feature on the model's predictions, we employ SHapley Additive
exPlanation analysis (SHAP) \citep{lundberg_unified_2017, rajbahadur_impact_2022}. We calculate the SHAP values for each project and then aggregate the results to gain a holistic understanding of the impact of a feature across all the studied projects.

\section{Maintainer Response Latency}
\label{sec:maintainers}
In this section, we aim to understand if we can accurately predict the first response latency of maintainers following the submission of a PR. Then, we want to identify and discuss the impact of the most important features in accurately predicting the first response latency of maintainers across the studied projects.

\subsection{RQ1: \rqi}
\label{sec:maintainers_rqi}

\Cref{tab:maintainer_models_aucroc} and \Cref{tab:maintainer_models_aucpr} compare the performance of various models for predicting the first response latency of maintainers in terms of the AUC-ROC and AUC-PR metrics, respectively. The best-performing models according to the Scott-Knott ESD test are highlighted in bold. It is worth noting that multiple models may be identified as best-performing when the difference in their performance is not statistically significant. From the tables, we find that the CB model consistently outperforms all other models in every project in both AUC-ROC and AUC-PR. Compared to the baseline (i.e., a dummy classifier that always returns the majority class), the CB model achieves considerable improvements, with increases ranging from 16\% to 45\% in AUC-ROC and from 20\% to 118\% in AUC-PR across different projects. The precision and recall scores of the CB model for different projects can be found in Appendix.

\begin{table*}
    \centering
    \caption{AUC-ROC of different models for predicting the first response latency of maintainers across the studied projects.}
    \label{tab:maintainer_models_aucroc}
    \begin{threeparttable}
        \begin{tabular}{@{}l|cccccccc@{}}
            \toprule
            \textbf{Project} & \textbf{CB}           & \textbf{KNN} & \textbf{LR}           & \textbf{NB}  & \textbf{NN}  & \textbf{RF}  & \textbf{SVM} \\
            \midrule
            Odoo             & \textbf{0.65 (+31\%)} & 0.59 (+18\%) & 0.63 (+26\%)          & 0.62 (+24\%) & 0.64 (+28\%) & 0.60 (+20\%) & 0.63 (+26\%) \\
            Kubernetes       & \textbf{0.63 (+25\%)} & 0.56 (+12\%) & 0.62 (+24\%)          & 0.60 (+19\%) & 0.61 (+22\%) & 0.56 (+12\%) & 0.58 (+16\%) \\
            Elasticsearch    & \textbf{0.67 (+33\%)} & 0.58 (+15\%) & 0.63 (+25\%)          & 0.61 (+22\%) & 0.64 (+29\%) & 0.56 (+12\%) & 0.57 (+15\%) \\
            PyTorch          & \textbf{0.64 (+29\%)} & 0.57 (+14\%) & 0.61 (+23\%)          & 0.62 (+24\%) & 0.62 (+25\%) & 0.57 (+14\%) & 0.59 (+18\%) \\
            Rust             & \textbf{0.59 (+19\%)} & 0.53 (+6\%)  & \textbf{0.59 (+18\%)} & 0.58 (+15\%) & 0.57 (+15\%) & 0.52 (+4\%)  & 0.53 (+7\%)  \\
            DefinitelyTyped  & \textbf{0.58 (+16\%)} & 0.54 (+9\%)  & \textbf{0.58 (+16\%)} & 0.55 (+10\%) & 0.56 (+11\%) & 0.55 (+10\%) & 0.57 (+13\%) \\
            HomeAssistant    & \textbf{0.69 (+38\%)} & 0.59 (+19\%) & 0.63 (+26\%)          & 0.61 (+22\%) & 0.67 (+34\%) & 0.57 (+15\%) & 0.59 (+18\%) \\
            Ansible          & \textbf{0.61 (+23\%)} & 0.56 (+12\%) & 0.60 (+20\%)          & 0.57 (+15\%) & 0.60 (+20\%) & 0.57 (+14\%) & 0.60 (+20\%) \\
            CockroachDB      & \textbf{0.67 (+35\%)} & 0.58 (+15\%) & 0.66 (+32\%)          & 0.63 (+25\%) & 0.64 (+27\%) & 0.58 (+15\%) & 0.57 (+14\%) \\
            Swift            & \textbf{0.63 (+27\%)} & 0.55 (+10\%) & 0.61 (+21\%)          & 0.59 (+19\%) & 0.62 (+23\%) & 0.54 (+7\%)  & 0.55 (+9\%)  \\
            Flutter          & \textbf{0.71 (+42\%)} & 0.61 (+23\%) & 0.67 (+35\%)          & 0.64 (+29\%) & 0.67 (+35\%) & 0.61 (+22\%) & 0.64 (+28\%) \\
            Spark            & \textbf{0.65 (+30\%)} & 0.57 (+13\%) & 0.61 (+22\%)          & 0.59 (+18\%) & 0.62 (+24\%) & 0.56 (+12\%) & 0.57 (+14\%) \\
            Python           & \textbf{0.62 (+24\%)} & 0.55 (+10\%) & 0.60 (+20\%)          & 0.57 (+15\%) & 0.60 (+19\%) & 0.56 (+12\%) & 0.59 (+18\%) \\
            Sentry           & \textbf{0.72 (+45\%)} & 0.60 (+19\%) & 0.66 (+31\%)          & 0.68 (+36\%) & 0.69 (+38\%) & 0.57 (+14\%) & 0.61 (+21\%) \\
            PaddlePaddle     & \textbf{0.62 (+24\%)} & 0.56 (+11\%) & 0.59 (+18\%)          & 0.60 (+19\%) & 0.60 (+21\%) & 0.55 (+11\%) & 0.58 (+15\%) \\
            Godot            & \textbf{0.62 (+24\%)} & 0.55 (+11\%) & \textbf{0.62 (+24\%)} & 0.60 (+20\%) & 0.60 (+21\%) & 0.54 (+9\%)  & 0.57 (+13\%) \\
            Rails            & \textbf{0.64 (+28\%)} & 0.57 (+13\%) & \textbf{0.64 (+27\%)} & 0.60 (+20\%) & 0.63 (+25\%) & 0.53 (+6\%)  & 0.57 (+15\%) \\
            Grafana          & \textbf{0.69 (+38\%)} & 0.59 (+19\%) & 0.65 (+31\%)          & 0.65 (+31\%) & 0.67 (+35\%) & 0.58 (+15\%) & 0.60 (+20\%) \\
            ClickHouse       & \textbf{0.63 (+25\%)} & 0.55 (+10\%) & 0.57 (+14\%)          & 0.58 (+15\%) & 0.60 (+20\%) & 0.56 (+11\%) & 0.57 (+14\%) \\
            Symfony          & \textbf{0.62 (+24\%)} & 0.54 (+8\%)  & \textbf{0.62 (+23\%)} & 0.60 (+20\%) & 0.60 (+19\%) & 0.52 (+3\%)  & 0.54 (+9\%)  \\
            \midrule
            Average          & \textbf{0.64 (+29\%)} & 0.57 (+13\%) & 0.62 (+24\%)          & 0.60 (+21\%) & 0.62 (+25\%) & 0.56 (+12\%) & 0.58 (+16\%) \\
            \bottomrule
        \end{tabular}
        \begin{tablenotes}
            \item Values in parentheses show the percentage improvement compared to the baseline.
        \end{tablenotes}
    \end{threeparttable}
\end{table*}

\begin{table*}
    \centering
    \caption{AUC-PR of different models for predicting the first response latency of maintainers across the studied projects.}
    \label{tab:maintainer_models_aucpr}
    \begin{threeparttable}
        \begin{tabular}{@{}l|cccccccc@{}}
            \toprule
            \textbf{Project} & \textbf{CB}            & \textbf{KNN} & \textbf{LR}           & \textbf{NB}  & \textbf{NN}   & \textbf{RF}  & \textbf{SVM} \\
            \midrule
            Odoo             & \textbf{0.48 (+45\%)}  & 0.40 (+21\%) & 0.44 (+34\%)          & 0.43 (+30\%) & 0.46 (+39\%)  & 0.40 (+19\%) & 0.45 (+34\%) \\
            Kubernetes       & \textbf{0.42 (+34\%)}  & 0.37 (+11\%) & 0.42 (+35\%)          & 0.40 (+25\%) & 0.41 (+28\%)  & 0.36 (+10\%) & 0.39 (+21\%) \\
            Elasticsearch    & \textbf{0.44 (+58\%)}  & 0.37 (+17\%) & 0.41 (+44\%)          & 0.39 (+37\%) & 0.42 (+49\%)  & 0.37 (+15\%) & 0.38 (+25\%) \\
            PyTorch          & \textbf{0.44 (+43\%)}  & 0.37 (+15\%) & 0.41 (+33\%)          & 0.41 (+32\%) & 0.42 (+36\%)  & 0.37 (+13\%) & 0.40 (+28\%) \\
            Rust             & \textbf{0.38 (+30\%)}  & 0.35 (+6\%)  & 0.37 (+25\%)          & 0.37 (+21\%) & 0.37 (+21\%)  & 0.34 (+4\%)  & 0.35 (+12\%) \\
            DefinitelyTyped  & \textbf{0.41 (+20\%)}  & 0.36 (+9\%)  & 0.40 (+21\%)          & 0.37 (+10\%) & 0.39 (+15\%)  & 0.36 (+8\%)  & 0.39 (+16\%) \\
            HomeAssistant    & \textbf{0.42 (+79\%)}  & 0.37 (+25\%) & 0.38 (+45\%)          & 0.38 (+35\%) & 0.41 (+67\%)  & 0.36 (+22\%) & 0.37 (+39\%) \\
            Ansible          & \textbf{0.43 (+33\%)}  & 0.38 (+13\%) & 0.41 (+26\%)          & 0.39 (+17\%) & 0.42 (+27\%)  & 0.37 (+12\%) & 0.42 (+27\%) \\
            CockroachDB      & \textbf{0.42 (+67\%)}  & 0.37 (+19\%) & 0.42 (+62\%)          & 0.39 (+44\%) & 0.41 (+48\%)  & 0.37 (+19\%) & 0.38 (+35\%) \\
            Swift            & \textbf{0.40 (+50\%)}  & 0.35 (+13\%) & 0.38 (+36\%)          & 0.37 (+24\%) & 0.38 (+41\%)  & 0.35 (+9\%)  & 0.35 (+20\%) \\
            Flutter          & \textbf{0.47 (+76\%)}  & 0.40 (+32\%) & 0.44 (+60\%)          & 0.41 (+47\%) & 0.45 (+63\%)  & 0.39 (+29\%) & 0.43 (+52\%) \\
            Spark            & \textbf{0.42 (+52\%)}  & 0.36 (+17\%) & 0.39 (+31\%)          & 0.38 (+23\%) & 0.40 (+39\%)  & 0.36 (+13\%) & 0.37 (+24\%) \\
            Python           & \textbf{0.41 (+30\%)}  & 0.36 (+10\%) & 0.40 (+26\%)          & 0.38 (+19\%) & 0.40 (+24\%)  & 0.37 (+12\%) & 0.40 (+23\%) \\
            Sentry           & \textbf{0.45 (+118\%)} & 0.37 (+31\%) & 0.41 (+83\%)          & 0.40 (+71\%) & 0.42 (+105\%) & 0.37 (+26\%) & 0.38 (+57\%) \\
            PaddlePaddle     & \textbf{0.42 (+37\%)}  & 0.37 (+15\%) & 0.39 (+29\%)          & 0.39 (+28\%) & 0.40 (+31\%)  & 0.36 (+10\%) & 0.39 (+25\%) \\
            Godot            & \textbf{0.41 (+35\%)}  & 0.36 (+12\%) & 0.40 (+31\%)          & 0.39 (+24\%) & 0.40 (+29\%)  & 0.35 (+8\%)  & 0.37 (+18\%) \\
            Rails            & \textbf{0.40 (+43\%)}  & 0.36 (+13\%) & \textbf{0.40 (+42\%)} & 0.38 (+25\%) & 0.39 (+34\%)  & 0.34 (+8\%)  & 0.36 (+21\%) \\
            Grafana          & \textbf{0.45 (+82\%)}  & 0.38 (+26\%) & 0.42 (+66\%)          & 0.41 (+57\%) & 0.43 (+72\%)  & 0.37 (+20\%) & 0.39 (+44\%) \\
            ClickHouse       & \textbf{0.41 (+43\%)}  & 0.36 (+12\%) & 0.38 (+23\%)          & 0.38 (+26\%) & 0.40 (+35\%)  & 0.36 (+11\%) & 0.38 (+27\%) \\
            Symfony          & \textbf{0.39 (+39\%)}  & 0.35 (+10\%) & 0.38 (+38\%)          & 0.38 (+27\%) & 0.37 (+28\%)  & 0.34 (+5\%)  & 0.35 (+19\%) \\
            \midrule
            Average          & \textbf{0.42 (+51\%)}  & 0.37 (+16\%) & 0.40 (+39\%)          & 0.39 (+31\%) & 0.41 (+41\%)  & 0.36 (+14\%) & 0.38 (+28\%) \\
            \bottomrule
        \end{tabular}
        \begin{tablenotes}
            \item Values in parentheses show the percentage improvement compared to the baseline.
        \end{tablenotes}
    \end{threeparttable}
\end{table*}

To understand the factors contributing to the misclassification of the first response latency of maintainers, we manually examined the misclassified PRs by the CB model. The most important reason is that the predictions are based on the data available up until the time of PR submission. However, there are various post-submission factors that affect how long it takes for maintainers to respond. For example, if a PR fails Continuous Integration (CI) tests, maintainers usually wait for the errors to be fixed before starting the review \citep{khatoonabadi_wasted_2023, li_are_2022}. Furthermore, the presence of bot-generated responses is known to prolong the first response latency of maintainers \citep{hasan_understanding_2023}.

\begin{tcolorbox}
    The CB model can predict the first response latency of maintainers with an average improvement of 29\% in AUC-ROC and 51\% in AUC-PR compared to a dummy classifier that always returns the majority class.
\end{tcolorbox}

\subsection{RQ2: \rqii}
\begin{figure*}
    \centering
    \includegraphics[width=\linewidth]{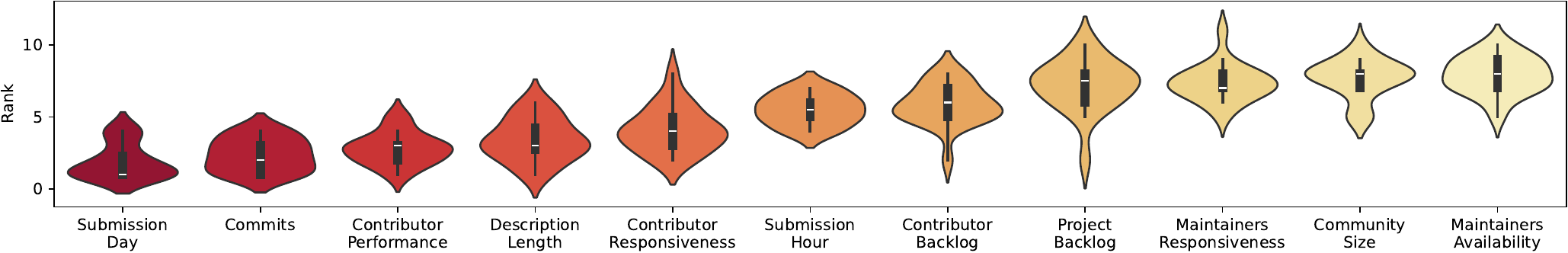}
    \caption{Ranking of the importance of different features for predicting the first response latency of maintainers across the studied projects. Darker colors indicate higher importance.}
    \label{fig:importances_maintainers}
\end{figure*}

To conduct this analysis, we use the CB models due to their superior performance. \Cref{fig:importances_maintainers} overviews the rankings of different features based on their importance in accurately predicting the first response latency of maintainers across the studied projects. From the figure, we observe that \textit{Submission Day}, \textit{Commits}, \textit{Contributor Performance}, \textit{Description Length}, and \textit{Contributor Responsiveness} are the top five most important features. This observation implies that the characteristics of PRs and contributors have a greater influence on how quickly maintainers provide their first response compared to project characteristics. We were surprised by this observation, as we expected that at least historical maintainer responsiveness to be a key predictor of future response times.

\begin{figure}
    \centering
    \includegraphics[width=\linewidth]{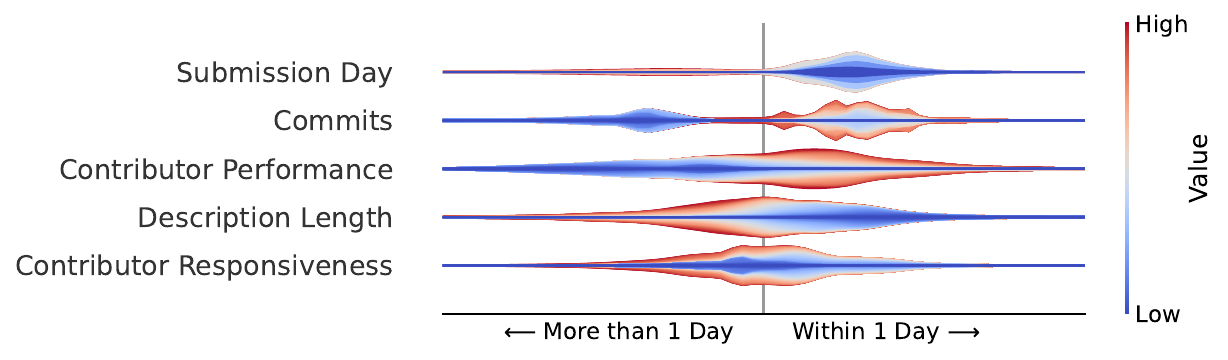}
    \caption{Impact of the top 5 most important features on the prediction of the first response latency of maintainers across the studied projects. Wider violins indicate higher density and more frequent values.}
    \label{fig:impacts_maintainers}
\end{figure}

\Cref{fig:impacts_maintainers} illustrates the impact of the top five most important features on the probability of receiving the first maintainer response on the same day of submitting a PR. We observe that PRs submitted earlier in the week, containing an average or slightly above-average number of commits at submission, and with more concise descriptions are more likely to get faster responses. Similarly, contributors with a higher acceptance rate and a history of timely responses in the project tend to obtain quicker responses. However, it is concerning that inexperienced contributors are prone to encounter delays in receiving feedback. This lack of timely responsiveness from maintainers is frequently cited as a key reason why contributors, especially novice or casual contributors, may abandon their PRs \citep{khatoonabadi_wasted_2023, li_are_2022, wang_why_2019} and even cease further contributing to the project \citep{assavakamhaenghan_does_2023, steinmacher_why_2013}.

\begin{tcolorbox}
    The \textit{Submission Day}, \textit{Commits}, \textit{Contributor Performance}, \textit{Description Length}, and \textit{Contributor Responsiveness} have the most influence on the first response latency of maintainers.
\end{tcolorbox}

\section{Contributor Response Latency}
\label{sec:contributors}
In this section, we aim to understand if we can accurately predict the first response latency of the contributor of a PR after receiving the first response from the maintainers. Then, we want to identify and discuss the impact of the most important features in accurately predicting the first response latency of contributors across the studied projects.

\subsection{RQ3: \rqiii}
\Cref{tab:contributor_models_aucroc} and \Cref{tab:contributor_models_aucpr} compare the performance of various models for predicting the first response latency of contributors in terms of the AUC-ROC and AUC-PR metrics, respectively. The best-performing models according to the Scott-Knott ESD test are highlighted in bold. It is worth noting that multiple models may be identified as best-performing when the difference in their performance is not statistically significant. From the tables, we find that similar to the first response latency of maintainers (see \Cref{sec:maintainers}), the CB model continues to demonstrate superior performance in both AUC-ROC and AUC-PR across all projects. Compared to the baseline (i.e., a dummy classifier that always returns the majority class), the CB model achieves significant improvements, with increases ranging from 24\% to 50\% in AUC-ROC and from 39\% to 153\% in AUC-PR across different projects. The precision and recall scores of the CB model for different projects can be found in Appendix.

To understand the factors contributing to the misclassification of the first response latency of contributors, we manually examined the misclassified PRs by the CB model. We find that the quality of feedback and the extent of requested changes also influence how long it takes for contributors to respond after the feedback. This observation aligns with prior findings in the literature that emphasize the importance of quality review comments from the maintainers \citep{li_opportunities_2022, kononenko_investigating_2015}.

\begin{table*}
    \centering
    \caption{AUC-ROC of different models for predicting the first response latency of contributors across the studied projects.}
    \label{tab:contributor_models_aucroc}
    \begin{threeparttable}
        \begin{tabular}{@{}l|cccccccc@{}}
            \toprule
            \textbf{Project} & \textbf{CB}           & \textbf{KNN} & \textbf{LR}           & \textbf{NB} & \textbf{NN}   & \textbf{RF}  & \textbf{SVM} \\
            \midrule
            Odoo             & \textbf{0.66 (+32\%)} & 0.55 (+11\%) & 0.64 (+28\%)          & 0.62 (+24\%) & 0.62 (+24\%) & 0.57 (+13\%) & 0.59 (+19\%) \\
            Kubernetes       & \textbf{0.68 (+37\%)} & 0.58 (+16\%) & 0.68 (+35\%)          & 0.64 (+27\%) & 0.65 (+30\%) & 0.59 (+17\%) & 0.61 (+21\%) \\
            Elasticsearch    & \textbf{0.74 (+48\%)} & 0.59 (+18\%) & 0.71 (+41\%)          & 0.67 (+35\%) & 0.69 (+38\%) & 0.58 (+16\%) & 0.60 (+21\%) \\
            PyTorch          & \textbf{0.68 (+35\%)} & 0.58 (+15\%) & 0.67 (+33\%)          & 0.64 (+29\%) & 0.64 (+28\%) & 0.58 (+16\%) & 0.60 (+20\%) \\
            Rust             & \textbf{0.64 (+28\%)} & 0.56 (+12\%) & \textbf{0.64 (+28\%)} & 0.62 (+23\%) & 0.61 (+21\%) & 0.56 (+12\%) & 0.55 (+11\%) \\
            DefinitelyTyped  & \textbf{0.62 (+24\%)} & 0.55 (+10\%) & 0.61 (+21\%)          & 0.59 (+18\%) & 0.60 (+21\%) & 0.54 (+8\%)  & 0.56 (+11\%) \\
            HomeAssistant    & \textbf{0.75 (+49\%)} & 0.62 (+23\%) & 0.72 (+44\%)          & 0.68 (+36\%) & 0.71 (+43\%) & 0.58 (+15\%) & 0.58 (+17\%) \\
            Ansible          & \textbf{0.66 (+32\%)} & 0.58 (+17\%) & \textbf{0.66 (+33\%)} & 0.61 (+22\%) & 0.64 (+27\%) & 0.57 (+15\%) & 0.61 (+22\%) \\
            CockroachDB      & \textbf{0.72 (+44\%)} & 0.58 (+15\%) & 0.71 (+41\%)          & 0.66 (+32\%) & 0.66 (+31\%) & 0.57 (+14\%) & 0.58 (+16\%) \\
            Swift            & \textbf{0.72 (+44\%)} & 0.58 (+16\%) & 0.69 (+38\%)          & 0.68 (+35\%) & 0.67 (+34\%) & 0.56 (+12\%) & 0.59 (+19\%) \\
            Flutter          & \textbf{0.72 (+44\%)} & 0.59 (+17\%) & 0.69 (+37\%)          & 0.62 (+23\%) & 0.65 (+30\%) & 0.56 (+13\%) & 0.60 (+20\%) \\
            Spark            & \textbf{0.73 (+45\%)} & 0.59 (+18\%) & 0.70 (+40\%)          & 0.69 (+38\%) & 0.68 (+36\%) & 0.57 (+14\%) & 0.61 (+21\%) \\
            Python           & \textbf{0.68 (+35\%)} & 0.58 (+15\%) & 0.67 (+33\%)          & 0.66 (+31\%) & 0.65 (+31\%) & 0.54 (+9\%)  & 0.58 (+16\%) \\
            Sentry           & \textbf{0.74 (+49\%)} & 0.58 (+17\%) & 0.69 (+38\%)          & 0.66 (+33\%) & 0.69 (+38\%) & 0.58 (+16\%) & 0.60 (+20\%) \\
            PaddlePaddle     & \textbf{0.67 (+35\%)} & 0.55 (+10\%) & 0.64 (+28\%)          & 0.62 (+24\%) & 0.61 (+22\%) & 0.55 (+10\%) & 0.56 (+12\%) \\
            Godot            & \textbf{0.67 (+35\%)} & 0.59 (+18\%) & \textbf{0.68 (+35\%)} & 0.66 (+32\%) & 0.66 (+31\%) & 0.56 (+12\%) & 0.61 (+23\%) \\
            Rails            & \textbf{0.68 (+36\%)} & 0.57 (+13\%) & 0.67 (+33\%)          & 0.62 (+25\%) & 0.62 (+24\%) & 0.55 (+10\%) & 0.58 (+15\%) \\
            Grafana          & \textbf{0.75 (+50\%)} & 0.60 (+21\%) & 0.71 (+42\%)          & 0.70 (+40\%) & 0.71 (+41\%) & 0.59 (+17\%) & 0.60 (+20\%) \\
            ClickHouse       & \textbf{0.70 (+41\%)} & 0.57 (+14\%) & 0.69 (+38\%)          & 0.64 (+28\%) & 0.67 (+34\%) & 0.57 (+13\%) & 0.59 (+19\%) \\
            Symfony          & \textbf{0.68 (+36\%)} & 0.57 (+15\%) & \textbf{0.68 (+35\%)} & 0.66 (+31\%) & 0.64 (+29\%) & 0.55 (+11\%) & 0.59 (+18\%) \\
            \midrule
            Average          & \textbf{0.69 (+39\%)} & 0.58 (+16\%) & 0.68 (+35\%)          & 0.65 (+29\%) & 0.65 (+31\%) & 0.57 (+13\%) & 0.59 (+18\%) \\
            \bottomrule
        \end{tabular}
        \begin{tablenotes}
            \item Values in parentheses show the percentage improvement compared to the baseline.
        \end{tablenotes}
    \end{threeparttable}
\end{table*}

\begin{table*}
    \centering
    \caption{AUC-PR of different models for predicting the first response latency of contributors across the studied projects.}
    \label{tab:contributor_models_aucpr}
    \begin{threeparttable}
        \begin{tabular}{@{}l|cccccccc@{}}
            \toprule
            \textbf{Project} & \textbf{CB}            & \textbf{KNN} & \textbf{LR}            & \textbf{NB}  & \textbf{NN}  & \textbf{RF}  & \textbf{SVM} \\
            \midrule
            Odoo             & \textbf{0.44 (+58\%)}  & 0.36 (+12\%) & 0.42 (+49\%)           & 0.40 (+35\%) & 0.41 (+39\%) & 0.37 (+15\%) & 0.39 (+34\%) \\
            Kubernetes       & \textbf{0.45 (+62\%)}  & 0.37 (+19\%) & 0.44 (+60\%)           & 0.41 (+40\%) & 0.43 (+48\%) & 0.38 (+20\%) & 0.40 (+37\%) \\
            Elasticsearch    & \textbf{0.46 (+142\%)} & 0.37 (+28\%) & 0.43 (+112\%)          & 0.40 (+72\%) & 0.42 (+97\%) & 0.37 (+32\%) & 0.38 (+58\%) \\
            PyTorch          & \textbf{0.45 (+70\%)}  & 0.37 (+19\%) & 0.43 (+60\%)           & 0.41 (+45\%) & 0.42 (+55\%) & 0.37 (+20\%) & 0.40 (+39\%) \\
            Rust             & \textbf{0.41 (+46\%)}  & 0.36 (+14\%) & \textbf{0.41 (+52\%)}  & 0.39 (+35\%) & 0.39 (+36\%) & 0.36 (+13\%) & 0.36 (+22\%) \\
            DefinitelyTyped  & \textbf{0.40 (+39\%)}  & 0.36 (+13\%) & 0.39 (+31\%)           & 0.38 (+23\%) & 0.39 (+34\%) & 0.35 (+9\%)  & 0.37 (+20\%) \\
            HomeAssistant    & \textbf{0.44 (+112\%)} & 0.37 (+36\%) & 0.43 (+102\%)          & 0.41 (+73\%) & 0.42 (+93\%) & 0.37 (+29\%) & 0.38 (+48\%) \\
            Ansible          & \textbf{0.45 (+51\%)}  & 0.38 (+21\%) & \textbf{0.45 (+50\%)}  & 0.41 (+32\%) & 0.42 (+40\%) & 0.37 (+16\%) & 0.41 (+34\%) \\
            CockroachDB      & \textbf{0.43 (+95\%)}  & 0.36 (+21\%) & 0.42 (+96\%)           & 0.39 (+52\%) & 0.39 (+61\%) & 0.36 (+25\%) & 0.37 (+38\%) \\
            Swift            & \textbf{0.43 (+108\%)} & 0.36 (+24\%) & 0.41 (+100\%)          & 0.39 (+69\%) & 0.40 (+72\%) & 0.36 (+31\%) & 0.37 (+54\%) \\
            Flutter          & \textbf{0.44 (+141\%)} & 0.37 (+30\%) & 0.42 (+124\%)          & 0.38 (+50\%) & 0.40 (+68\%) & 0.36 (+42\%) & 0.38 (+50\%) \\
            Spark            & \textbf{0.45 (+106\%)} & 0.37 (+27\%) & 0.44 (+104\%)          & 0.42 (+75\%) & 0.42 (+77\%) & 0.37 (+24\%) & 0.39 (+59\%) \\
            Python           & \textbf{0.42 (+63\%)}  & 0.36 (+20\%) & \textbf{0.42 (+65\%)}  & 0.40 (+51\%) & 0.41 (+56\%) & 0.35 (+12\%) & 0.38 (+37\%) \\
            Sentry           & \textbf{0.46 (+153\%)} & 0.36 (+21\%) & 0.42 (+121\%)          & 0.38 (+64\%) & 0.41 (+87\%) & 0.38 (+45\%) & 0.37 (+48\%) \\
            PaddlePaddle     & \textbf{0.41 (+76\%)}  & 0.35 (+14\%) & 0.39 (+66\%)           & 0.38 (+47\%) & 0.38 (+50\%) & 0.35 (+19\%) & 0.36 (+30\%) \\
            Godot            & \textbf{0.42 (+63\%)}  & 0.37 (+26\%) & \textbf{0.42 (+65\%)}  & 0.40 (+52\%) & 0.41 (+62\%) & 0.36 (+20\%) & 0.39 (+46\%) \\
            Rails            & \textbf{0.41 (+76\%)}  & 0.36 (+20\%) & \textbf{0.41 (+74\%)}  & 0.38 (+47\%) & 0.38 (+50\%) & 0.36 (+22\%) & 0.37 (+40\%) \\
            Grafana          & \textbf{0.44 (+120\%)} & 0.37 (+30\%) & 0.43 (+115\%)          & 0.40 (+79\%) & 0.41 (+91\%) & 0.37 (+29\%) & 0.37 (+51\%) \\
            ClickHouse       & \textbf{0.46 (+114\%)} & 0.37 (+34\%) & \textbf{0.45 (+115\%)} & 0.40 (+52\%) & 0.43 (+93\%) & 0.37 (+34\%) & 0.39 (+66\%) \\
            Symfony          & \textbf{0.42 (+78\%)}  & 0.36 (+24\%) & \textbf{0.41 (+77\%)}  & 0.39 (+50\%) & 0.40 (+57\%) & 0.36 (+25\%) & 0.38 (+46\%) \\
            \midrule
            Average          & \textbf{0.43 (+89\%)}  & 0.36 (+23\%) & 0.42 (+82\%)           & 0.40 (+52\%) & 0.41 (+63\%) & 0.36 (+24\%) & 0.38 (+43\%) \\
            \bottomrule
        \end{tabular}
        \begin{tablenotes}
            \item Values in parentheses show the percentage improvement compared to the baseline.
        \end{tablenotes}
    \end{threeparttable}
\end{table*}

\begin{tcolorbox}
    The CB model can predict the first response latency of contributors with an average improvement of 39\% in AUC-ROC and 89\% in AUC-PR compared to a dummy classifier that always returns the majority class.
\end{tcolorbox}

\subsection{RQ4: \rqiv}
\begin{figure*}
    \centering
    \includegraphics[width=\linewidth]{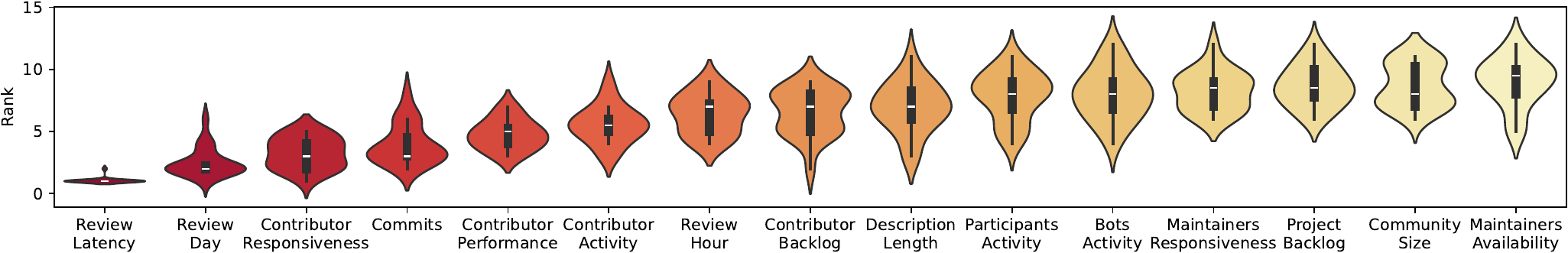}
    \caption{Ranking of the importance of different features for predicting the first response latency of contributors across the studied projects. Darker colors indicate higher importance.}
    \label{fig:importances_contributors}
\end{figure*}

To conduct this analysis, we use the CB models due to their superior performance. \Cref{fig:importances_contributors} overviews the rankings of different features based on their importance in accurately predicting the first response latency of contributors across the studied projects. From the figure, we observe that \textit{Review Latency}, \textit{Review Day}, \textit{Contributor Responsiveness}, \textit{Commits}, and \textit{Contributor Performance} are the top five most important features. This observation highlights the great influence of the characteristics of the review process and contributors on how quickly the contributors first respond.

\begin{figure}
    \centering
    \includegraphics[width=\linewidth]{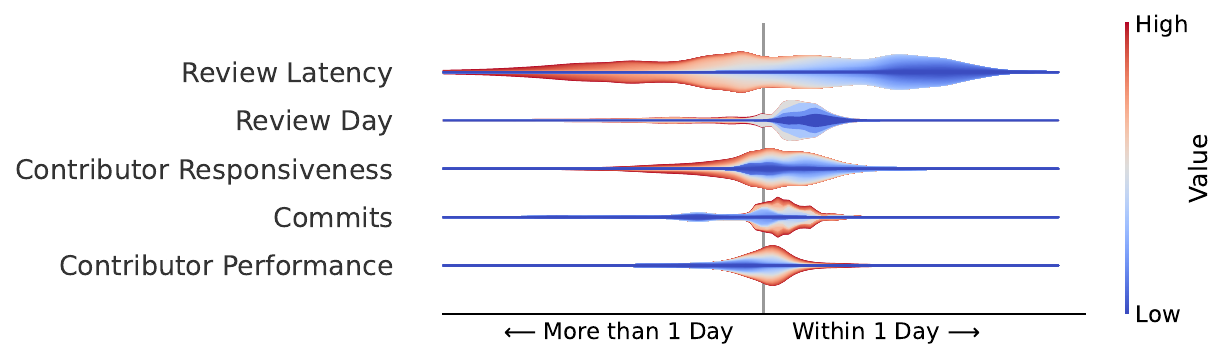}
    \caption{Impact of the top 5 most important features on the prediction of the first response latency of contributors across the studied projects. Wider violins indicate higher density and more frequent values.}
    \label{fig:impacts_contributors}
\end{figure}

\Cref{fig:impacts_contributors} illustrates the impact of the top five most important features on the probability of contributors replying on the same day they receive the first maintainers response. We observe that lower latency in the first maintainer response correlates with a quicker subsequent response from the contributor. In other words, contributors tend to reply late if they have experienced delayed responses, leading to a cascade effect in the review process. Notably, the first response latency of maintainers not only affects the contributor responsiveness but is also known to directly impact the duration \citep{hasan_understanding_2023, zhang_pull_2022} and the outcome \citep{zhang_pull_2023, khatoonabadi_wasted_2023, li_are_2022} of the PR, as well as the likelihood of future contributions by the contributor to the project \citep{assavakamhaenghan_does_2023, hasan_understanding_2023}. Furthermore, we find contributors of PRs that received the first response of maintainers earlier in the week, and contributors of PRs containing an average or slightly above-average number of commits till the first response of maintainers are more likely to provide faster responses. Similarly, contributors with a history of timely responses in the project and with a higher acceptance rate tend to give quicker responses.

\begin{tcolorbox}
    The \textit{Review Latency}, \textit{Review Day}, \textit{Contributor Responsiveness}, \textit{Commits}, and \textit{Contributor Performance} have the most influence on the first response latency of contributors.
\end{tcolorbox}

\section{Cross-Project Setting}
\label{sec:cross-project}
Building accurate predictive models for new projects is often challenging due to the limited historical data available. However, one way to overcome this challenge is through cross-project prediction, which enables such projects to leverage the insights and patterns observed in older, well-established projects. To evaluate the effectiveness of this approach, for each studied project, we train a CB model (as it has shown superior performance in our previous analyses) on all other projects and then test on that project.

\Cref{tab:generic_models_performance} compares the performance of various models in predicting the first response latency of maintainers and contributors in a cross-project setting in terms of the AUC-ROC and AUC-PR metrics, respectively. The precision and recall scores of the CB model for different projects can also be found in Appendix. We observe that the models for predicting the maintainers response latency achieve improvements ranging from 22\% to 58\% in AUC-ROC and from 28\% to 122\% in AUC-PR. The models for predicting the contributors response latency also demonstrate improvements ranging from 26\% to 56\% in AUC-ROC and from 35\% to 149\% in AUC-PR. The results indicate that our approach can be effective for predicting both the first response latency of maintainers and the first response latency of contributors in a cross-project setting.

Furthermore, we find that \textit{Submission Day}, \textit{Commits}, \textit{Contributor Performance}, \textit{Maintainers Responsiveness}, and \textit{Contributor Responsiveness} are the major predictors of the first response latency of maintainers in our cross-project models (see Appendix). Compared to our project-specific analysis of maintainers' response latencies (RQ2), \textit{Maintainers Responsiveness} now replaces \textit{Description Length} as the 4th most important feature, with \textit{Description Length} now ranked 6th and \textit{Maintainers Responsiveness} previously ranked 9th. The increased importance of \textit{Maintainers Responsiveness} (i.e., median first response latency of the maintainers over the last 3 months) in our cross-project models may be attributed to the higher variation in this feature's values when using data from other projects for training. In contrast, project-specific models, which use data from only a given project, typically exhibit a lower variation of this feature. We also observe that \textit{Review Latency}, \textit{Review Day}, \textit{Contributor Responsiveness}, \textit{Commits}, and \textit{Contributor Activity} are the major predictors of the first response latency of contributors in our cross-project models (see Appendix). Compared to our project-specific analysis of contributors' response latencies (RQ4), \textit{Contributor Activity} ranks as the 5th most important feature in our cross-project analysis, whereas it was ranked 6th in the project-specific analysis. Similarly, \textit{Contributor Performance}, previously ranked 5th in the project-specific analysis, is now ranked 6th. This indicates that the key predictors of contributors' response latencies are quite similar in both our project-specific and cross-project models.

\begin{table}
    \centering
    \caption{Performance of the models for predicting the first response latency of maintainers and contributors in a cross-project setting.}
    \label{tab:generic_models_performance}
    \resizebox{\linewidth}{!}{
        \begin{threeparttable}
            \begin{tabular}{@{}l|cc|cc@{}}
                \toprule
                                 & \multicolumn{2}{c}{\textbf{Maintainers}} & \multicolumn{2}{|c}{\textbf{Contributors}} \\
                \textbf{Project} & \textbf{AUC-ROC} & \textbf{AUC-PR}       & \textbf{AUC-ROC} & \textbf{AUC-PR}         \\
                \midrule
                Odoo             & 0.64 (+28\%)     & 0.46 (+37\%)          & 0.66 (+32\%)     & 0.44 (+57\%)            \\
                Kubernetes       & 0.68 (+36\%)     & 0.45 (+50\%)          & 0.70 (+40\%)     & 0.46 (+73\%)            \\
                Elasticsearch    & 0.67 (+35\%)     & 0.44 (+61\%)          & 0.74 (+49\%)     & 0.46 (+133\%)           \\
                PyTorch          & 0.66 (+33\%)     & 0.45 (+50\%)          & 0.70 (+40\%)     & 0.46 (+71\%)            \\
                Rust             & 0.64 (+29\%)     & 0.40 (+48\%)          & 0.67 (+35\%)     & 0.43 (+65\%)            \\
                DefinitelyTyped  & 0.61 (+22\%)     & 0.43 (+28\%)          & 0.63 (+26\%)     & 0.41 (+35\%)            \\
                HomeAssistant    & 0.65 (+30\%)     & 0.39 (+55\%)          & 0.74 (+48\%)     & 0.44 (+106\%)           \\
                Ansible          & 0.64 (+28\%)     & 0.45 (+38\%)          & 0.69 (+39\%)     & 0.47 (+69\%)            \\
                CockroachDB      & 0.74 (+49\%)     & 0.45 (+100\%)         & 0.74 (+48\%)     & 0.44 (+121\%)           \\
                Swift            & 0.68 (+35\%)     & 0.41 (+63\%)          & 0.73 (+46\%)     & 0.43 (+111\%)           \\
                Flutter          & 0.79 (+58\%)     & 0.50 (+122\%)         & 0.78 (+56\%)     & 0.46 (+149\%)           \\
                Spark            & 0.67 (+34\%)     & 0.44 (+61\%)          & 0.73 (+46\%)     & 0.46 (+111\%)           \\
                Python           & 0.61 (+22\%)     & 0.41 (+29\%)          & 0.69 (+39\%)     & 0.43 (+66\%)            \\
                Sentry           & 0.71 (+42\%)     & 0.44 (+107\%)         & 0.74 (+47\%)     & 0.44 (+140\%)           \\
                PaddlePaddle     & 0.65 (+31\%)     & 0.43 (+43\%)          & 0.67 (+33\%)     & 0.40 (+72\%)            \\
                Godot            & 0.63 (+27\%)     & 0.41 (+35\%)          & 0.68 (+37\%)     & 0.42 (+69\%)            \\
                Rails            & 0.66 (+33\%)     & 0.43 (+64\%)          & 0.71 (+43\%)     & 0.44 (+101\%)           \\
                Grafana          & 0.70 (+40\%)     & 0.45 (+77\%)          & 0.76 (+52\%)     & 0.45 (+136\%)           \\
                ClickHouse       & 0.62 (+24\%)     & 0.40 (+35\%)          & 0.72 (+44\%)     & 0.48 (+122\%)           \\
                Symfony          & 0.64 (+28\%)     & 0.41 (+53\%)          & 0.70 (+40\%)     & 0.42 (+85\%)            \\
                \midrule
                Average          & 0.67 (+33\%)     & 0.43 (+58\%)          & 0.71 (+42\%)     & 0.44 (+95\%)            \\
                \bottomrule
            \end{tabular}
            \begin{tablenotes}
                \item Values in parentheses show the percentage improvement compared to the baseline.
            \end{tablenotes}
        \end{threeparttable}
    }
\end{table}

\section{Implications}
\label{sec:implications}
In the following, we discuss key guidelines for maintainers, contributors, and researchers to help facilitate the PR review process.

\subsection{Implications for Contributors}
Our findings on the response latency of maintainers (RQ2) suggest several strategies for contributors to improve their chances of receiving faster feedback from maintainers. First, the timing of PR submissions is crucial. We recommend that contributors submit their PRs on weekdays, ideally earlier in the week, to take advantage of the potentially higher availability of maintainers during this period. In addition, the complexity of PRs is important. We recommend that contributors ensure their PRs are substantial enough to warrant attention but not so large as to deter maintainers from engaging promptly. Furthermore, the quality of PR descriptions matters. We recommend that contributors strive to write succinct descriptions to make it easier for maintainers to understand and evaluate the proposed changes. Lastly, our findings suggest that the track record of contributors also appears to play a significant role. Therefore, we recommend that contributors maintain a history of timely interactions and successful contributions to increase the likelihood of quick feedback. Over time, such consistent performance can build a reputation that potentially encourages faster and more favorable responses from maintainers.

\subsection{Implications for Maintainers}
Our findings on the response latency of contributors (RQ4) suggest several strategies for maintainers to facilitate the PR review process. First, the first response latency of maintainers is crucial, as delayed responses are very likely to trigger a cascading effect that adversely impacts the entire review process. We recommend that maintainers strive to provide timely responses to set a positive tone and sustain the momentum of the review process. In addition, the timing of review comments is important. We recommend that maintainers submit their feedback on weekdays, especially earlier in the week. Furthermore, our findings suggest that maintainers seem to favor contributors with a history of successful contributions within the project. It is important for maintainers to be mindful of the potential demotivating effects of delayed responses on less experienced contributors. Being more responsive toward novice or casual contributors can encourage their continued participation and foster a more inclusive and collaborative environment.

\subsection{Implications for Researchers}
A promising direction for future work is the development of a comprehensive approach based on our proposed models. This approach would not only predict anticipated waiting times but also explain the specific reasons for predicted delays using techniques such as LIME (Local Interpretable Model-agnostic Explanations \citep{ribeiro_why_2016}). Most importantly, the approach should extend beyond diagnostics to intervention, by proactively providing customized recommendations to minimize potential waiting times. This feature would enhance the practicality and relevance of the approach in real-world scenarios. A crucial aspect of this work is the empirical evaluation of the approach. This evaluation should assess the impact of the approach on enhancing collaboration and productivity among maintainers and contributors. Key metrics for this assessment can include the decrease in response times, the increase in PR success rates, and the overall satisfaction of all parties involved.

\section{Limitations}
\label{sec:limitations}
In this section, we discuss threats to the validity of our study.

\header{Internal Validity.} The first threat relates to the completeness of our features. To mitigate this threat, we consulted the literature on pull-based development \citep{zhang_pull_2022, zhang_pull_2023} and also drew from our previous experience studying PR abandonment \citep{khatoonabadi_wasted_2023, khatoonabadi_understanding_2023}. Nonetheless, there may be other features that we did not consider or that are challenging to quantify, such as code quality, feedback quality, and PR urgency, which potentially have a stronger association with the response latencies. The second threat relates to the choice of classifiers. To mitigate this threat, we employed seven different classifiers commonly used in the software engineering literature. CatBoost is also acclaimed for its effectiveness with multi-class imbalanced datasets, making it particularly suitable for our study. The third threat relates to the completeness of our maintainer identification approach. To mitigate this threat, we considered not only developers who have previously merged a PR or closed someone else's PR but also those who performed any of the privileged PR-related events identified by \citet{bock_automatic_2023} as maintainers. The fourth threat relates to the completeness of our bot detection approach. To mitigate this threat, we manually examined actors with high activity levels or fast response times and added them to our bot list accordingly. The fifth threat relates to our reliance on GitHub usernames to differentiate between different actors in PRs. In cases where a contributor uses multiple accounts, our measurement of contributor features and the first response latency of contributors can be impacted. Similarly, when a maintainer uses multiple accounts, our approach identifies only the account that has previously performed a privileged event as a maintainer, which can affect our measurements for project features and the first response latency of maintainers. To further improve the robustness of our approach, we recommend that future work investigate techniques for correcting such developer identity errors by cross-referencing multiple identity information beyond just usernames \citep{amreen_alfaa_2020}. The sixth threat relates to all timestamps being standardized to UTC as provided by the GitHub API. This may not accurately reflect the working hours of developers in different time zones, potentially affecting the interpretation of temporal features like \textit{Submission Hour} and \textit{Review Hour}. To address this limitation, we recommend that future work explore methods to accurately extract the local time zones of developers at each specific event throughout a project's history.

\header{External Validity.} Our study is based on 20 large and popular open-source projects on GitHub. Although we believe these projects are more likely to benefit from our approach due to their higher activity levels, we recognize that they cannot represent the entire open-source ecosystem, especially smaller projects. In other words, the studied projects may not represent other open-source projects with different sizes, maturity, popularity, workload, dynamics, social structures, and development practices. Still, we evaluated our approach in a cross-project setting to demonstrate its effectiveness across different projects. Future research can replicate our approach using a more diverse selection of projects.

\section{Related Work}
\label{sec:related_work}
In the following, we first overview the studies on the first response latency of PRs before moving on to the studies on the review duration and outcome of PRs. Finally, we review the studies on the reasons, consequences, and solutions to PR abandonment.

\header{Studies on First Response Latency.} \citet{hasan_understanding_2023} is the first to conduct an exploratory study on the time-to-first-response of bots and humans in PRs. They found that first responses in PRs are often generated by bots. They also observed that complex PRs with lengthy descriptions and inexperienced contributors with less communicative attitudes tend to experience longer delays in receiving the first human response. \citet{kudrjavets_mining_2022} reported the waiting time from the proposal of code changes until the first response as nonproductive time that negatively affects code velocity. \textit{While these studies focus on understanding the first response latency in PRs, our work specifically predicts the first response latency of maintainers following the submission of a PR. Besides, we also predict the first response latency of the contributor of a PR after receiving the first response from the maintainers.}

\header{Studies on Review Duration and Outcome.} The literature has extensively studied the influence of various technical and social factors \citep{soares_acceptance_2015, yu_determinants_2016, kononenko_studying_2018, pinto_who_2018, zou_how_2019, lenarduzzi_does_2021}, as well as various personal and demographic factors \citep{rastogi_biases_2016, terrell_gender_2017, rastogi_relationship_2018, furtado_how_2021, iyer_effects_2021, nadri_insights_2021, nadri_relationship_2022} on the review duration and outcome of PRs. Recently, \citet{zhang_pull_2022} conducted a large-scale empirical study to understand how a range of factors, identified through a systematic literature review, can explain the latency of PRs under different scenarios. They found that the description length is the most influential factor when PRs are submitted. When closing PRs, using CI tools, or when the contributor and the integrator differ, the presence of comments is the most influential factor. When comments are present, the latency of the first comment is the most influential.

\citet{zhang_pull_2023} also conducted a similar comprehensive empirical study to investigate how a range of factors, identified through a systematic literature review, can explain the decision of PRs under different scenarios. Most notably, they found that the area hotness of PR is influential only in the early stage of project development and becomes less influential as projects mature. \textit{While these studies investigated what factors influence the review duration and outcome of PRs, our work focuses on predicting the first response latency of maintainers and contributors.}

\header{Studies on PR Abandonment.} PR abandonment leads to a significant loss of time and effort for both contributors and maintainers, while simultaneously adding to the complexity of project management and maintenance tasks. \citet{li_are_2022} found that the primary reasons for PR abandonment are a lack of responsiveness from maintainers and a lack of time or interest from contributors. Furthermore, they reported that abandoned PRs lead to clutter in the lists of PRs, wasted review resources, additional work required for proper closure, delayed integration of interdependent PRs, duplication of PRs, disruption of project milestones, and ultimately the potential to leave a negative impression on the community.

\citet{khatoonabadi_wasted_2023} also studied the influence of various factors characterizing PRs, contributors, review processes, and projects on PR abandonment. They found that complex PRs, PRs from novice contributors, and PRs with long discussions are more likely to get abandoned. They also observed that the obstacles faced by the contributors and the hurdles imposed by the maintainers during the review process of PRs are the most frequent reasons for PR abandonment.

To deal with abandoned PRs, Stale bot was released in 2017 to automatically triage abandoned issues and PRs and is adopted by many open-source projects \citep{github_stale_2023, keepers_stale_2023, wessel_should_2019}. However, studies \citep{wessel_dont_2021, liu_understanding_2020, wessel_inconvenient_2020, farah_exploratory_2022, rahman_towards_2022} have mentioned that Stale bot introduces noise and friction for both the contributors and the maintainers. \citet{khatoonabadi_understanding_2023} conducted an empirical study to understand the helpfulness of Stale bot in the context of pull-based development. They found that Stale bot can help projects deal with a backlog of unresolved PRs and also improve the review process of PRs. However, the adoption of Stale bot can negatively affect the contributors (especially novice or casual contributors) in a project. \textit{While these studies investigated the reasons and consequences of PR abandonment and the helpfulness of Stale bot as a common solution to abandoned PRs, our work focuses on predicting the first response latency of maintainers and contributors to facilitate collaboration between maintainers and contributors during the review process of PRs.}

\section{Conclusion}
\label{sec:conclusion}
The objective of this paper was to develop the first machine-learning approach for predicting the latency of both maintainers' first response following a PR submission and the contributor's first response after receiving the first maintainer feedback. We curated a dataset of 20 large and popular open-source projects on GitHub and extracted 21 features characterizing projects, contributors, PRs, and review processes. The analyses demonstrated the effectiveness of our approach in predicting the first response latency of maintainers and contributors both in project-specific and cross-project settings. Furthermore, the findings highlighted the significant influence of the timing of submissions and feedback, the complexity of changes, and the track record of contributors on the response latencies. By providing estimated waiting times, our approach can help open-source projects facilitate collaboration between their maintainers and contributors by enabling them to anticipate and address potential delays proactively during PR review process.

\bibliographystyle{IEEEtranN}
\bibliography{references}

\newpage

\begin{IEEEbiography}[{\includegraphics[width=1in,height=1.25in,clip,keepaspectratio]{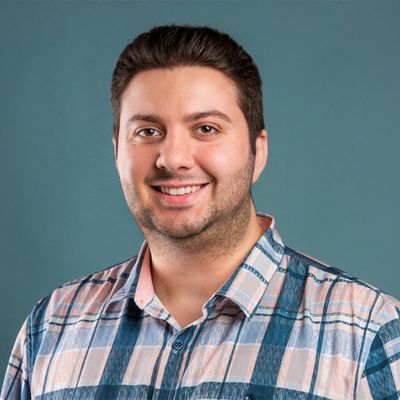}}]{SayedHassan Khatoonabadi}
is a Postdoctoral Fellow (NSERC CREATE SE4AI) in the Department of Computer Science and Software Engineering at Concordia University, Canada. He received his Ph.D. from Concordia University, Canada. His research interests include developer experience, SE4AI, AI4SE, mining software repositories, and empirical software engineering. You can find more about him at \url{https://linkedin.com/in/khatoonabadi}.
\end{IEEEbiography}

\begin{IEEEbiography}[{\includegraphics[width=1in,height=1.25in,clip,keepaspectratio]{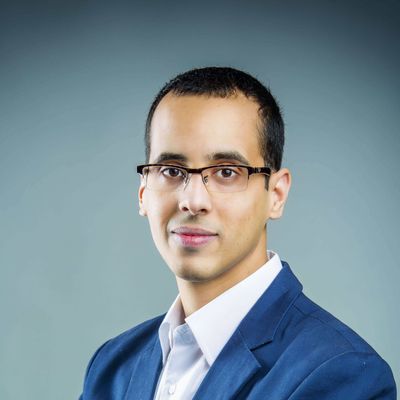}}]{Ahmad Abdellatif}
is an Assistant Professor in the Department of Electrical and Software Engineering at the University of Calgary, Canada. He received his Ph.D. from Concordia University, Canada. His research interests include software chatbots, mining software repositories, SE4AI, and empirical software engineering. You can find more about him at \url{https://aabdllatif.github.io}.
\end{IEEEbiography}

\begin{IEEEbiography}[{\includegraphics[width=1in,height=1.25in,clip,keepaspectratio]{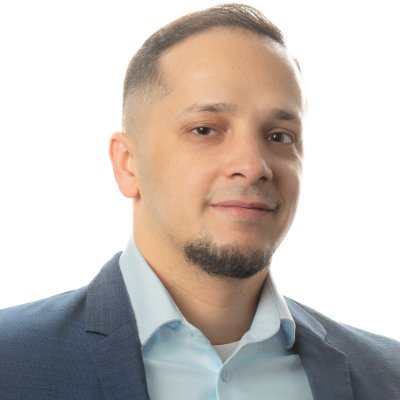}}]{Diego Elias Costa}
is an Assistant Professor in the Department of Computer Science and Software Engineering at Concordia University, Canada. Before that, he was an Assistant Professor in the Department of Computer Science at UQAM, Canada. He received his Ph.D. from Heidelberg University, Germany. His research interests include SE4AI, dependency management, performance testing, and software engineering bots. You can find more about him at \url{https://realiselab.github.io/teamInfo/diego}.
\end{IEEEbiography}

\begin{IEEEbiography}[{\includegraphics[width=1in,height=1.25in,clip,keepaspectratio]{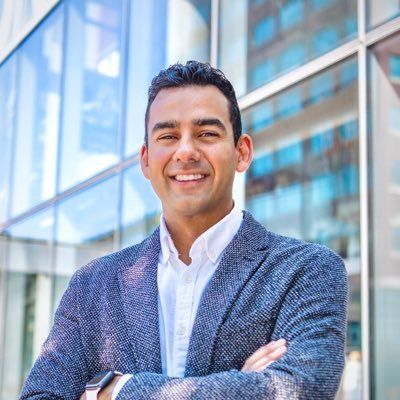}}]{Emad Shihab}
is Associate Dean of Research and Innovation and Full Professor in the Gina Cody School of Engineering and Computer Science at Concordia University, Canada. He is a member of the Royal Society of Canada and a Senior Member of IEEE. He received his Ph.D. from Queen's University, Canada. His research interests include software engineering, mining software repositories, and software analytics.  You can find more about him at \url{https://das.encs.concordia.ca/members/emad-shihab}.
\end{IEEEbiography}

\vfill

\onecolumn
\appendices
\section*{Appendix}
\label{sec:appendix}

\begin{table*}[h]
    \centering
    \caption{Distribution of the first response latency of maintainers and contributors across the studied projects.}
    \resizebox{\textwidth}{!}{%
        \begin{tabular}{@{}l|cccc|cccc@{}}
            \toprule
                             & \multicolumn{4}{c}{\textbf{Maintainers}}                                                       & \multicolumn{4}{c}{\textbf{Contributors}}                                                      \\
            \textbf{Project} & \textbf{Within 1 Day} & \textbf{1 Day to 1 Week} & \textbf{More than 1 Week} & \textbf{\# PRs} & \textbf{Within 1 Day} & \textbf{1 Day to 1 Week} & \textbf{More than 1 Week} & \textbf{\# PRs} \\
            \midrule
            Odoo             & 47.3\%                & 21.4\%                   & 31.3\%                    & 38,105          & 72.2\%                & 17.8\%                   & 10.1\%                    & 21,498          \\
            Kubernetes       & 65.5\%                & 21.4\%                   & 13.1\%                    & 59,172          & 72.6\%                & 17.8\%                   & 9.6\%                     & 41,729          \\
            Elasticsearch    & 73.0\%                & 19.4\%                   & 7.6\%                     & 40,646          & 82.8\%                & 13.3\%                   & 4.0\%                     & 37,132          \\
            PyTorch          & 66.2\%                & 24.3\%                   & 9.5\%                     & 43,787          & 73.0\%                & 19.5\%                   & 7.5\%                     & 29,806          \\
            Rust             & 75.9\%                & 17.5\%                   & 6.6\%                     & 45,943          & 75.5\%                & 17.6\%                   & 6.9\%                     & 29,085          \\
            DefinitelyTyped  & 40.1\%                & 42.6\%                   & 17.3\%                    & 42,617          & 72.7\%                & 19.3\%                   & 8.0\%                     & 13,585          \\
            HomeAssistant    & 80.6\%                & 13.3\%                   & 6.1\%                     & 44,333          & 82.5\%                & 11.9\%                   & 5.6\%                     & 23,942          \\
            Ansible          & 53.2\%                & 24.1\%                   & 22.8\%                    & 36,521          & 69.0\%                & 17.1\%                   & 13.9\%                    & 17,086          \\
            CockroachDB      & 80.7\%                & 15.5\%                   & 3.8\%                     & 41,677          & 86.6\%                & 10.3\%                   & 3.2\%                     & 40,631          \\
            Swift            & 81.8\%                & 13.0\%                   & 5.2\%                     & 29,138          & 85.7\%                & 10.3\%                   & 4.0\%                     & 22,397          \\
            Flutter          & 77.1\%                & 15.6\%                   & 7.4\%                     & 24,503          & 84.1\%                & 11.8\%                   & 4.1\%                     & 20,629          \\
            Spark            & 71.3\%                & 18.8\%                   & 9.9\%                     & 29,999          & 80.2\%                & 14.9\%                   & 4.9\%                     & 24,041          \\
            Python           & 64.4\%                & 16.5\%                   & 19.2\%                    & 17,759          & 79.9\%                & 11.9\%                   & 8.1\%                     & 11,649          \\
            Sentry           & 83.8\%                & 13.4\%                   & 2.9\%                     & 33,020          & 86.2\%                & 11.3\%                   & 2.4\%                     & 31,616          \\
            PaddlePaddle     & 62.0\%                & 29.3\%                   & 8.7\%                     & 26,155          & 84.2\%                & 12.4\%                   & 3.4\%                     & 18,711          \\
            Godot            & 68.4\%                & 16.3\%                   & 15.3\%                    & 27,307          & 81.2\%                & 10.6\%                   & 8.2\%                     & 12,664          \\
            Rails            & 78.5\%                & 10.9\%                   & 10.6\%                    & 27,693          & 83.7\%                & 9.9\%                    & 6.4\%                     & 14,387          \\
            Grafana          & 75.6\%                & 17.8\%                   & 6.6\%                     & 23,116          & 84.2\%                & 11.8\%                   & 3.9\%                     & 18,436          \\
            ClickHouse       & 68.5\%                & 21.7\%                   & 9.8\%                     & 15,058          & 75.2\%                & 18.6\%                   & 6.2\%                     & 7,823           \\
            Symfony          & 77.6\%                & 14.9\%                   & 7.5\%                     & 25,582          & 82.8\%                & 11.6\%                   & 5.7\%                     & 15,034          \\
            \bottomrule
        \end{tabular}
    }
\end{table*}

\begin{table*}[h]
    \centering
    \caption{Precison and recall scores of CatBoost models for predicting the first response latency of maintainers and contributors across the studied projects.}
    \begin{threeparttable}
        \begin{tabular}{@{}l|cc|cc@{}}
            \toprule
                             & \multicolumn{2}{c}{\textbf{Maintainers}} & \multicolumn{2}{c}{\textbf{Contributors}} \\
            \textbf{Project} & \textbf{Precision} & \textbf{Recall}     & \textbf{Precision} & \textbf{Recall}      \\
            \midrule
            Odoo             & 0.48 (+208\%)      & 0.46 (+37\%)        & 0.46 (+92\%)       & 0.41 (+23\%)         \\
            Kubernetes       & 0.43 (+105\%)      & 0.40 (+20\%)        & 0.46 (+94\%)       & 0.42 (+25\%)         \\
            Elasticsearch    & 0.45 (+83\%)       & 0.39 (+18\%)        & 0.53 (+91\%)       & 0.41 (+22\%)         \\
            PyTorch          & 0.44 (+106\%)      & 0.40 (+21\%)        & 0.47 (+96\%)       & 0.40 (+21\%)         \\
            Rust             & 0.39 (+55\%)       & 0.36 (+7\%)         & 0.43 (+71\%)       & 0.37 (+12\%)         \\
            DefinitelyTyped  & 0.40 (+206\%)      & 0.38 (+15\%)        & 0.40 (+66\%)       & 0.38 (+13\%)         \\
            HomeAssistant    & 0.45 (+68\%)       & 0.38 (+13\%)        & 0.47 (+70\%)       & 0.39 (+17\%)         \\
            Ansible          & 0.45 (+162\%)      & 0.42 (+25\%)        & 0.45 (+99\%)       & 0.42 (+26\%)         \\
            CockroachDB      & 0.44 (+69\%)       & 0.39 (+17\%)        & 0.47 (+64\%)       & 0.40 (+19\%)         \\
            Swift            & 0.43 (+57\%)       & 0.37 (+12\%)        & 0.46 (+62\%)       & 0.40 (+20\%)         \\
            Flutter          & 0.47 (+86\%)       & 0.44 (+31\%)        & 0.47 (+71\%)       & 0.41 (+23\%)         \\
            Spark            & 0.45 (+85\%)       & 0.38 (+14\%)        & 0.49 (+82\%)       & 0.40 (+19\%)         \\
            Python           & 0.41 (+92\%)       & 0.40 (+19\%)        & 0.43 (+61\%)       & 0.38 (+15\%)         \\
            Sentry           & 0.48 (+70\%)       & 0.41 (+23\%)        & 0.54 (+87\%)       & 0.42 (+25\%)         \\
            PaddlePaddle     & 0.42 (+111\%)      & 0.39 (+17\%)        & 0.43 (+52\%)       & 0.38 (+13\%)         \\
            Godot            & 0.42 (+81\%)       & 0.38 (+14\%)        & 0.44 (+62\%)       & 0.40 (+19\%)         \\
            Rails            & 0.41 (+56\%)       & 0.37 (+10\%)        & 0.44 (+57\%)       & 0.37 (+11\%)         \\
            Grafana          & 0.48 (+88\%)       & 0.41 (+23\%)        & 0.45 (+58\%)       & 0.39 (+18\%)         \\
            ClickHouse       & 0.43 (+90\%)       & 0.40 (+19\%)        & 0.50 (+100\%)      & 0.42 (+25\%)         \\
            Symfony          & 0.40 (+53\%)       & 0.36 (+8\%)         & 0.43 (+59\%)       & 0.39 (+16\%)         \\
            \midrule
            Average          & 0.44 (+97\%)       & 0.39 (+18\%)        & 0.46 (+75\%)       & 0.40 (+19\%)         \\
            \bottomrule
        \end{tabular}
        \begin{tablenotes}
            \item Values in parentheses show the percentage improvement compared to the baseline.
        \end{tablenotes}
    \end{threeparttable}
\end{table*}

\begin{table*}[h]
    \centering
    \caption{Precison and recall scores of CatBoost models for predicting the first response latency of maintainers and contributors in a cross-project setting.}
    \begin{threeparttable}
        \begin{tabular}{@{}l|cc|cc@{}}
            \toprule
                             & \multicolumn{2}{c}{\textbf{Maintainers}} & \multicolumn{2}{c}{\textbf{Contributors}} \\
            \textbf{Project} & \textbf{Precision} & \textbf{Recall}     & \textbf{Precision} & \textbf{Recall}      \\
            \midrule
            Odoo             & 0.46 (+188\%)      & 0.41 (+24\%)        & 0.48 (+100\%)      & 0.37 (+12\%)         \\
            Kubernetes       & 0.46 (+109\%)      & 0.40 (+21\%)        & 0.49 (+104\%)      & 0.35 (+6\%)          \\
            Elasticsearch    & 0.53 (+121\%)      & 0.34 (+3\%)         & 0.60 (+114\%)      & 0.35 (+6\%)          \\
            PyTorch          & 0.50 (+127\%)      & 0.37 (+12\%)        & 0.56 (+133\%)      & 0.35 (+6\%)          \\
            Rust             & 0.43 (+72\%)       & 0.34 (+3\%)         & 0.47 (+88\%)       & 0.35 (+6\%)          \\
            DefinitelyTyped  & 0.44 (+238\%)      & 0.38 (+15\%)        & 0.42 (+75\%)       & 0.36 (+9\%)          \\
            HomeAssistant    & 0.35 (+30\%)       & 0.35 (+6\%)         & 0.51 (+89\%)       & 0.35 (+6\%)          \\
            Ansible          & 0.47 (+161\%)      & 0.40 (+21\%)        & 0.53 (+130\%)      & 0.38 (+15\%)         \\
            CockroachDB      & 0.45 (+67\%)       & 0.36 (+9\%)         & 0.53 (+83\%)       & 0.36 (+9\%)          \\
            Swift            & 0.49 (+81\%)       & 0.39 (+18\%)        & 0.41 (+41\%)       & 0.36 (+9\%)          \\
            Flutter          & 0.41 (+58\%)       & 0.34 (+3\%)         & 0.49 (+75\%)       & 0.36 (+9\%)          \\
            Spark            & 0.52 (+117\%)      & 0.35 (+6\%)         & 0.54 (+100\%)      & 0.35 (+6\%)          \\
            Python           & 0.48 (+129\%)      & 0.34 (+3\%)         & 0.45 (+67\%)       & 0.36 (+9\%)          \\
            Sentry           & 0.81 (+189\%)      & 0.34 (+3\%)         & 0.62 (+114\%)      & 0.34 (+3\%)          \\
            PaddlePaddle     & 0.45 (+114\%)      & 0.36 (+9\%)         & 0.38 (+36\%)       & 0.35 (+6\%)          \\
            Godot            & 0.40 (+74\%)       & 0.34 (+3\%)         & 0.44 (+63\%)       & 0.35 (+6\%)          \\
            Rails            & 0.66 (+154\%)      & 0.35 (+6\%)         & 0.53 (+89\%)       & 0.36 (+9\%)          \\
            Grafana          & 0.51 (+104\%)      & 0.34 (+3\%)         & 0.57 (+104\%)      & 0.34 (+3\%)          \\
            ClickHouse       & 0.33 (+43\%)       & 0.34 (+3\%)         & 0.63 (+152\%)      & 0.35 (+6\%)          \\
            Symfony          & 0.55 (+112\%)      & 0.34 (+3\%)         & 0.44 (+57\%)       & 0.34 (+3\%)          \\
            \midrule
            Average          & 0.49 (+114\%)      & 0.36 (+9\%)         & 0.50 (+91\%)       & 0.35 (+7\%)          \\
            \bottomrule
        \end{tabular}
        \begin{tablenotes}
            \item Values in parentheses show the percentage improvement compared to the baseline.
        \end{tablenotes}
    \end{threeparttable}
\end{table*}

\begin{figure*}[h]
    \includegraphics[width=\linewidth]{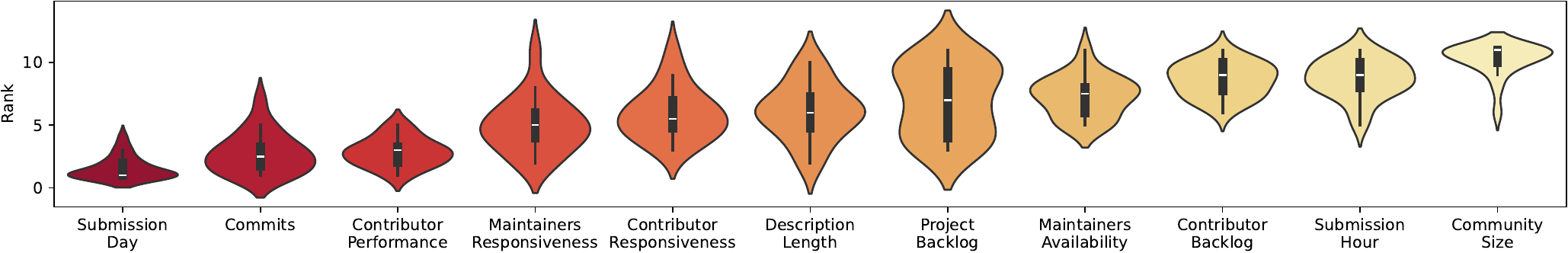}
    \caption{Ranking of the importance of different features for predicting the first response latency of maintainers in a cross-project scenario.}
\end{figure*}

\begin{figure*}[h]
    \includegraphics[width=\linewidth]{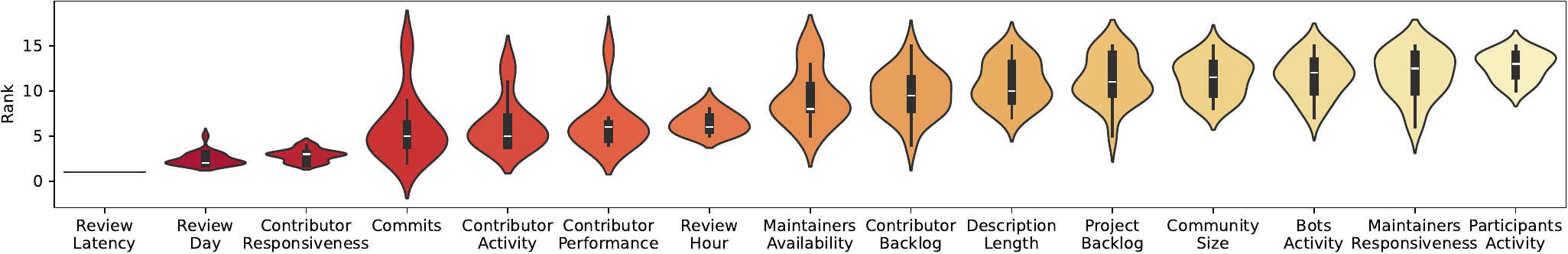}
    \caption{Ranking of the importance of different features for predicting the first response latency of contributors in a cross-project scenario.}
\end{figure*}

\end{document}